\documentclass[aps,prl,floats,twocolumn,10pt,floatfix,superscriptaddress]{revtex4-1}
\usepackage{graphicx}
\usepackage{lipsum, babel}
\usepackage{amssymb}
\usepackage{amsmath}
\newcommand\numberthis{\addtocounter{equation}{1}\tag{\theequation}}
\usepackage{color}
\usepackage[dvipsnames]{xcolor}
\usepackage{slantsc}
\usepackage{float}
\graphicspath{{./FIG_FINALI/}}
\usepackage{bbold}
\usepackage[caption=false]{subfig}
\usepackage{array}
\usepackage{braket}
\usepackage{scrextend}
\usepackage{hyperref}
\hypersetup{backref,pdfpagemode=FullScreen,colorlinks=true,allcolors=blue}
\usepackage{leftidx}

\begin{document}
\title{Changes in Polarization Dictate Necessary Approximations for Modeling Electronic De-excitation Intensity: an Application to X-ray Emission}
\author{Subhayan Roychoudhury}
\email{subhayan@lbl.gov}
\affiliation{The Molecular Foundry, Lawrence Berkeley National Laboratory, Berkeley CA 94720, USA}
\author{Leonardo A. Cunha}
\email{leonardo.cunha@berkeley.edu}
\affiliation{Department of Chemistry, University of California Berkeley, CA 94720 , USA}
\affiliation{Chemical Sciences Division, Lawrence Berkeley National Laboratory, Berkeley CA 94720 USA}
\author{Martin Head-Gordon}
\email{mhg@cchem.berkeley.edu}
\affiliation{Department of Chemistry, University of California Berkeley, CA 94720 , USA}
\affiliation{Chemical Sciences Division, Lawrence Berkeley National Laboratory, Berkeley CA 94720 USA}
\author{David Prendergast}
\email{dgprendergast@lbl.gov}
\affiliation{The Molecular Foundry, Lawrence Berkeley National Laboratory, Berkeley CA 94720, USA}

\begin{abstract}
We systematically investigate the underlying relations among different levels of approximation for simulating electronic de-excitations, with a focus on modeling X-ray emission spectroscopy (XES). Using Fermi's golden rule and explicit modeling of the initial, core-excited state and the final, valence-hole state, we show that XES can be accurately modeled by using orbital optimization for the various final states within a Slater-determinant framework. However, in this paper, we introduce a much cheaper approach reliant only on a single self-consistent field for all the final states, and show that it is typically sufficient. Further approximations reveal that these fundamentally many-body transitions can be reasonably approximated by projections of ground state orbitals, but that the ground state alone is insufficient. Furthermore, except in cases where the core-ionization induces negligible changes in polarization, the widely used linear-response approaches within the adiabatic approximation will have difficulty in accurately modeling de-excitation to the core level. Therefore, change in the net dipole moment of the valence electrons can serve as a metric for the validity of the linear-response approximation.
\end{abstract}

\maketitle

Density Functional Theory (DFT)~\cite{PhysRev.136.B864, PhysRev.140.A1133} has been the primary workhorse for electronic structure calculations over the last few decades~\cite{doi:10.1063/1.4869598, RevModPhys.87.897, C4CP90074J, doi:10.1098/rsta.2012.0488}, thanks to the accuracy it offers for modest computational expense. DFT-based first-principles calculations have not only helped in analyzing a plethora of experiments~\cite{besleys, Choudhary2021} performed on various materials, but have also contributed to efforts in the design and discovery of new functional materials~\cite{doi:10.1063/1.4812323, Curtarolo2013}. Although Kohn-Sham (KS) DFT, in its pristine form, gives us access only to ground state properties, multiple methods, leveraging different levels of approximations, have been developed for accessing excited state properties starting from the ground-state information provided by DFT~\cite{Hait2021, CULLEN201111, doi:10.1021/jp401323d, doi:10.1021/acs.jctc.0c00597, doi:10.1063/1.5018615, doi:10.1063/1.2977989, Roychoudhury2020, PhysRevLett.76.1212, doi:10.1142/9789812830586_0005, PhysRevLett.80.794, PhysRevB.62.4927}. These methods differ widely in terms of their accuracy, computational expense and the nature of the target systems. 

Computational schemes that rely on the KS framework for evaluating the energy and the oscillator strength of electronic excitation or de-excitation can be broadly categorized into two classes: (1) \textit{Excited state-specific DFT} (eDFT) methods assume that electronic excited states can be approximated by  a single Slater determinant (SD) obtained by a non-aufbau filling of the KS orbitals, with excited holes (electrons) represented by removing (adding) contributions to the electron density of particular orbitals below (above) the Fermi level; (2) Response-function based approaches, such as linear-response (LR) time-dependent DFT (TDDFT) or the Bethe-Salpeter Equation (BSE), operate by explicitly creating electron-hole pairs in a given reference determinant, with the final state represented by a linear combination of the SDs so obtained~\cite{RevModPhys.74.601}. Naturally, it is instructive to perform a detailed comparative assessment of these two distinct formalisms and to delineate, with proper explanations, the circumstances under which their results are likely to differ.

In this letter, we focus on electronic de-excitations with specific emphasis on calculating the oscillator strength. In particular, even though much of our discussions can be applied to any kind of electronic de-excitation, here we focus on simulating valence-to-core de-excitations, since this can be directly probed by X-ray emission spectroscopy (XES) experiments. 

Core-level spectroscopy, which involves excitation/de-excitation of electrons from/to core levels using photons of X-ray frequencies, is itself an active area of experimental and computational research and has consistently played a key role in the experimental investigation of electronic properties of a wide variety of functional materials~\cite{KOTANI19967, doi:10.1021/cr9900681, GrunertKlementiev+2020}. Therefore, efficient theoretical models~\cite{PhysRevB.83.115106, PhysRevB.25.5150, REHR2009548, doi:10.1021/acs.accounts.0c00171, doi:10.1021/acs.jctc.8b01046, PhysRevResearch.2.042003, PhysRevLett.118.096402, PhysRevB.95.155121, C9CP06592J, RevModPhys.83.705, PhysRevB.42.5459,PhysRevB.100.075149,doi:10.1021/acs.jpclett.9b03661} are always in demand for complementing these experimental efforts. On the other hand, from the perspective of a theorist, experimental X-ray spectroscopy results can serve as a paramount benchmark against which the efficacy of new theoretical approaches in electronic structure can be assessed~\cite{doi:10.1021/acs.jpclett.9b03661,PhysRevLett.118.096402}. Thus, a detailed comparison of the aforementioned theoretical frameworks in terms of valence-to-core de-excitations not only advances our understanding of electronic structure theory, but also serves the practical purpose of predicting and analyzing XES, which is a key component of materials-research. 

Since XES does not require any explicit description of the unoccupied orbitals, it can be simulated more reliably, compared to X-ray absorption spectroscopy (XAS), using a localized-basis calculation with a relatively smaller basis set. Additionally, unlike XAS, the LR-TDDFT based simulation of XES does not require any additional approximation of orbital-resolution, since all the solutions with negative eigenvalues correspond to de-excitation to the core level. This motivates us to use de-excitation (as opposed to excitation) as the focus of our investigation.

In non-resonant XES~\cite{doi:10.1021/cr9900681,Bergmann2009,https://doi.org/10.1002/jcc.26153,doi:10.1063/1.4977178}, a core-electron is ejected from a sample with the help of X-ray photons and as the core-excited system decays from an initial state $\Psi_I$ with energy $E_I$ to any possible final state $\Psi_F$ with energy $E_F$, the intensity
\begin{align}
    I(\omega) \propto \sum_F (E_I - E_F)^4 |M^{I,F}|^2 \delta(\omega+E_F-E_I),
\end{align}
of the emission is recorded as a function of its frequency $\omega$, where, denoting the transition operator by $\hat{O}$, the transition amplitude is given by $M^{I,F} = \braket{\Psi_F|\hat{O}|\Psi_I}$. Therefore $\ket{\Psi_{\rm{I}}}$, i.e., the state prior to the X-ray emission process, can be represented by the lowest energy state of the positively charged system in presence of a core-hole on the excited atom. In our calculations, this is approximated by a SD composed of KS orbitals obtained from a self-consistent field (SCF) KS DFT calculation, where the core-hole is created by imposing an additional constraint using the maximum overlap method (MOM)~\cite{doi:10.1021/jp801738f} such that, if there are $N$ electrons remaining in the system, then, denoting the creation operator for the $j$-th initial-state orbital $\tilde{\phi}_j$ by ${\tilde{a}}_j^\dagger$, we have

\begin{align}\label{I_SD}
     \ket{\Psi_{\rm{I}}} = \left(\prod_{j=1}^N{\tilde{a}}_j^\dagger\right)\ket{0}.
\end{align}

\begin{figure}
\centering
\includegraphics[width=0.5\textwidth]{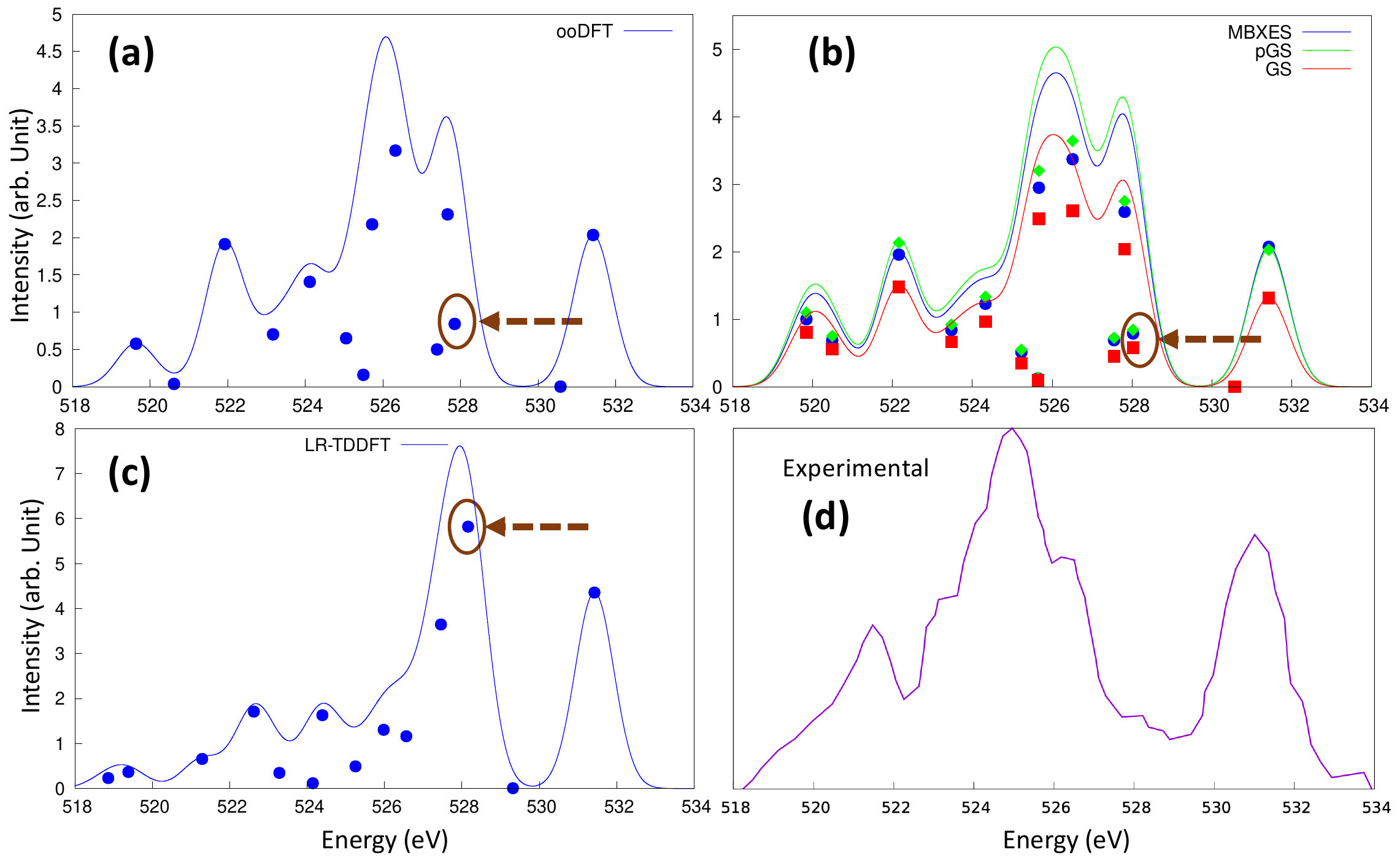}
\caption{O K-edge XES spectra of phenol calculated using various techniques, with energy (in eV) plotted along x-axis and intensity (in arbitrary units) plotted along y axis. Panel (a) shows the ooDFT results while panel (b) shows the MBXES, pGS, and GS results where the dipole matrix elements are calculated using Eq.~\ref{ESD_mat}, Eq.~\ref{MBXES_mat}, Eq.~\ref{pGS_Mat} and Eq.~\ref{GS_Mat}, respectively. The LR-TDDFT spectrum (Eq.~\ref{TDDFT_DipoleMat}) is shown in panel (c) while the experimental spectrum (adapted from~\cite{Yumatov1997}) is shown in panel (d). The 23-rd de-excitation, which is further analyzed in Fig.~\ref{OrbitalPlot} and S5, is shown by an arrow in panels (a)-(c).}\label{PhenolSpectra}
\end{figure}

As discussed below, within the eDFT framework, different approximations can be used to simulate the final states $\ket{\Psi_F}$, leading to different expressions for the transition amplitude. We illustrate these approximations by reference to Fig~\ref{PhenolSpectra}, which shows the O K-edge XES of a phenol molecule corresponding to these different approximations calculated with the Q-Chem code~\cite{doi:10.1063/5.0055522} using the $\omega$B97M-V xc-functional~\cite{doi:10.1063/1.4952647} and the aug-cc-pCVQZ basis-set. 

\textit{Fig.~\ref{PhenolSpectra}(a)} --- The orbital-optimized DFT (ooDFT) method employs a fully self-consistent procedure, similar to that used for obtaining $\ket{\Psi_I}$, such that a final state $\ket{\Psi_F}$ with filled core and a hole in the $f$-th valence orbital is approximated by the SD

\begin{align}\label{StateSpecific_SD}
     \ket{\Psi_F}=\ket{\Psi^{(f)}_{-f}} = {a}^{(f)}_f\left(\prod_{j=1}^N \left({a}^{(f)}_j\right)^{\dagger}\right) {a}_0^\dagger \ket{0}
\end{align}

where the subscript $0$ denotes the core orbital and $\left({a}^{(f)}_j\right)^{\dagger}$ is the creation operator for the $j$-th KS orbital $\phi^{(f)}_j$ corresponding to the SCF of a system with a hole in the $f$-th orbital, preserved, for example, using MOM. The resulting transition amplitude is then written as 
\begin{align*}\label{ESD_mat}
    M^F = M^{f}_{\rm{ooDFT}} = \braket{{\Psi^{(f)}_{-f}}|\hat{O}|\Psi_{\rm{I}}} &{} = \sum_{j=1}^N  \Xi^{(f)}_{j,f} \braket{\tilde{\phi}_{0}|\hat{o}|\tilde{\phi}_j}\numberthis,
\end{align*}
where we drop the $I$ superscript on $M$ as the initial state will be the same for all de-excitations in XES (i.e., the core-excited state). Here $\hat{o}$ denotes the single-particle transition operator, $\Xi^{(f)}_{j,f} = \braket{{\Psi^{(f)}_{-f}}|\Psi_{\mathrm{I}}^{+0-j}}$ is the determinantal overlap where  $\Psi_{\mathrm{I}}^{+0-j}=\tilde{a}_j\tilde{a}_0^\dagger\ket{\Psi_{\rm{I}}}$ [see the SI for a detailed derivation]. Note that, the transition moment is identical to that of the single-particle de-excitation of a non-interacting electron to the core level from an auxiliary orbital given by 
\begin{align}\label{AuxiliaryOrbital}
    \ket{\phi^{\rm{ooDFT}}_f} = \sum_{j=1}^N \Xi^{f}_{j,f} \ket{\tilde{\phi_j}} .
\end{align}

\textit{Fig.~\ref{PhenolSpectra}(b) [blue line]\footnote{In panels (b)-(c), a rigid shift has been applied to the spectra so that the highest-frequency point coincides with the energy $\left(E_{\rm{I}} - E_{\rm{GS}}^{+1}\right)$, where $E_{\rm{GS}}^{+1}$ is the total energy of the valence cationic ground state.}} --- Now, in order to avoid performing a separate SCF calculation for each final state, we propose the Many-Body X-ray Emission Spectroscopy (MBXES) approximation whereby the $F-$th final state, containing a hole in the $f-$th valence orbital is approximated as

\begin{align}\label{MBXES_SD}
     \ket{\Psi_F} = \ket{\Psi_{-f}} = {a}_f\left(\prod_{j=1}^N {a}_j^\dagger \right) {a}_0^\dagger \ket{0}
\end{align}
where ${a}_j^\dagger$ is the creation operator for the $j-$th KS orbital ($\phi_j$) of the ground-state SCF. Note that this approximation, which selects the orbitals constituting the SD of any final (core-filled) state from the same set of orbitals $\{\phi_j\}$, is equivalent to one that was proposed recently~\cite{PhysRevLett.118.096402,PhysRevB.97.205127,PhysRevB.100.075121} for simulating X-ray absorption and is essentially derived from the ooDFT approximation by neglecting the relaxation of electrons in response to the creation of a valence hole. Defining the relevant determinantal overlap as $\Xi_{j,f} = \braket{\Psi_{-f}|\Psi^{+0-j}_{\rm{I}}}$, the transition amplitude is then given by
\begin{align*}\label{MBXES_mat}
   M^{f}_{\rm{MBXES}} =  \braket{\Psi_{-f}|\hat{O}|\Psi_{\rm{I}}} &{}= \sum_{j=1}^N  \Xi_{j,f} \braket{\tilde{\phi}_0|\hat{o}|\tilde{\phi}_j}\numberthis,
\end{align*}
with  $\ket{\phi^{\rm{MBXES}}_f} = \sum_{j=1}^N \Xi_{j,f} \ket{\tilde{\phi_j}}$ representing the corresponding auxiliary orbital. 

\textit{Fig.~\ref{PhenolSpectra}(b) [green line]} Now, making the additional approximation $\Xi_{j,f}=\braket{\Psi_{-f}|\Psi_{\mathrm{I}}^{+0-j}} \approx - \braket{\tilde{\phi_j}|\phi_f} \equiv -\xi^*_{jf}$, i.e., approximating the overlap of the SDs with the negative conjugate of the overlap of their respective hole orbitals (see SI for more details), we can transition from a determinantal formalism to the \textit{projected Ground State} (pGS) treatment, where, ignoring the phase factor of $-1$, the transition moment is given by
\begin{align}\label{pGS_Mat}
    M^{f}_{\rm{pGS}} = \sum_{j=1}^N \xi^*_{jf}\braket{\tilde{\phi_0}|\hat{o}|\tilde{\phi}_j},
\end{align}
while $\ket{\phi^{\rm{pGS}}_f} = \sum_{j=1}^N \xi^*_{jf}\ket{\tilde{\phi}_j}$ gives the corresponding auxiliary orbital. 

\textit{Fig.~\ref{PhenolSpectra}(b) [red line]} Note that in all of the aforementioned approximations, separate sets of orbitals $\{\tilde{\phi}_j\}$ and $\{\phi_j\}$ (or $\{\phi_j^{(f)}\}$) are used for the initial and the final state, indicating separate SCFs for states with and without a core-hole. 
The final approximation within the eDFT framework neglects this effect by extending the sum in Eq.~\ref{pGS_Mat} to all orbitals and thereby approximating the transition moment entirely within a \textit{Ground State} (GS) treatment as 
\begin{align*}\label{GS_Mat}
    M^{f}_{\rm{GS}} =\sum_{j=1}^{\rm{all}} \xi^*_{jf}\braket{\tilde{\phi}_0|\hat{o}|\tilde{\phi}_j} = \braket{\phi_0|\hat{o}|\phi_f},\numberthis
\end{align*}
where the approximation that the core orbital remains unchanged, $\ket{\phi_0} \approx \ket{\tilde{\phi}_0}$, is validated by our calculations showing $\braket{\phi_0|\tilde{\phi}_0}\approx 1$. 

In summary, ooDFT (Eq.~\ref{ESD_mat}) allows for the relaxation of all the electrons consistent with the effect of the core-hole (valence-hole) in the initial (final) state. MBXES (Eq.~\ref{MBXES_mat}) uses an identical expression for $\ket{\Psi_{\rm{I}}}$, but neglects the effect of the valence hole, using frozen orbitals from a neutral ground-state calculation. pGS (Eq.~\ref{pGS_Mat}) takes into account the fact that the participating orbital should be different in presence and in absence of the core-hole, but it does not incorporate the relaxation of the other electrons explicitly. And the GS (Eq.~\ref{GS_Mat}) approximation ignores all effects of the core hole entirely. It is noteworthy that several previous studies~\cite{VU2020106287,PhysRevB.96.125136,doi:10.1021/jp511931y,JMSRR2018} have modeled XES using the ground state projected density-of-states (pDOS), a simplified version of the aforementioned GS formalism.

\textit{Fig.~\ref{PhenolSpectra}(c)} --- In contrast with the eDFT formalism, the response-function based approaches like LR-TDDFT and BSE, in their most widely used form,  operate by constructing and diagonalizing a 2-particle effective Hamiltonian ($H^{2\rm{p}}$) written in the electron-hole pair basis~\cite{RevModPhys.74.601}. This encodes the effects of electron-hole interactions with the help of a kernel which, in its exact form, should be non-local in space and time. 
Every eigenstate of $H^{2\rm{p}}$ can be associated with an approximate final-state $\ket{\Psi_F^{\rm{Resp}}}$ of the system, expressible as a linear combination of SDs (analogous to the configurations-interaction singles approach) each of which represents a transition of an electron from an occupied to an empty orbital (core orbital for XES) of the reference SD $\Psi_{\rm{I}}$:
\begin{align}\label{TDDFT_FinalSTate}
    \ket{\Psi_F^{\rm{Resp}}} = \sum_{j=1}^N \gamma^{F}_j {\tilde{a}}_j{a}_0^\dagger \ket{\Psi_{\rm{I}}},
\end{align}
such that the transition amplitude
\begin{align}\label{TDDFT_DipoleMat}
    M^{F}_{\rm{Resp}} = \braket{\Psi_F^{\rm{Resp}}|\hat{O}|\Psi_{\rm{I}}} = \sum_{j=1}^N \gamma^{F}_{j}\braket{\phi_{\rm{core}}|\hat{o}|\tilde{\phi}_j}.
\end{align}
is identical to that of a single-particle de-excitation from the auxiliary orbital $\ket{\phi^{\rm{Resp}}_F} = \sum_{j=1}^N \gamma^F_{j} \ket{\tilde{\phi}_j}$. 

Having introduced the relevant methods and demonstrated the corresponding spectra, it is now instructive to investigate the underlying connections among them. This is facilitated by the fact that all transition moments are now expressed as linear combinations of valence-to-core transitions from the occupied initial-state orbitals, which are identical for all methods. For example, a comparison of Eq.~\ref{pGS_Mat} and~\ref{GS_Mat} reveals that if a certain ground state orbital $\ket{\phi_i}$ has no overlap with the unoccupied subspace of the initial state, then the corresponding transition will have the same oscillator strength in the GS and the pGS picture. Since this does not guarantee an equality between $\Xi_{j,f}$ and $\xi_{j,f}$ $(\forall j \leq N)$, $M^f_{\rm{MBXES}}$ can, in principle still be different. However, as shown in the SI, from purely linear-algebraic considerations, it follows that if the valence occupied subspaces of the ground and the core-ionized (initial) state are identical, then $\xi_{j,f} = \pm \Xi_{j,f}$ $\forall f,j \leq N$; resulting in identical spectra for GS, pGS and MBXES, even if the GS valence orbitals differ from the initial state orbitals by unitary transformations. 

This leads to the following key insight: for systems in which the core-hole induces a negligible change in the valence density or polarization, a GS treatment should be comparable to MBXES. On the other hand, if the polarization resulting from the core-hole creation is appreciable (very likely), a GS calculation might prove to be inadequate and a determinantal treatment is recommended. 

Including the response-based approaches in the comparison, we note that the adiabatic approximation (which neglects the frequency-dependence of the kernel) is typically adequate for simulating valence excitations but might introduce larger errors in core-level excitations/de-excitations since they involve large exciton-binding energies. In order to take a more intuitive look into such possible inadequacy, we express the exact $F$-th final state as

 \begin{align*}\label{FSSD_op}
   \ket{\Psi_{F}^{\rm{exact}}} = &{} \left[\sum_{j=1}^N \alpha_j^F {\tilde{a}}_j{\tilde{a}}_0^\dagger \ket{\Psi_{\rm{I}}}\right] +\\ &{} \left[\sum_{l,m=1}^N\sum_{p=N+1}^{\rm{all}} \beta^F_{l,m,p} {\tilde{a}}_l {\tilde{a}}_m {\tilde{a}}_p^\dagger {\tilde{a}}_0^\dagger \ket{\Psi_{\rm{I}}}\right] + \hdots , \numberthis
\end{align*}
where the terms inside the first (second) set of square brackets represent the first (second)-order terms obtained by creating one (two) electron-hole pairs in $\ket{\Psi_{\rm{I}}}$. Note that only the first order terms can contribute to the transition-dipole matrix $M^F_{\rm{exact}}= \braket{\Psi_F^{\rm{exact}}|\hat{O}|\Psi_{\rm{I}}} = \sum_{j=1}^N \alpha_j^F \braket{\phi_0|\hat{o}|\tilde{\phi}_j}$, since for all other terms, the overlap with $\hat{O}\ket{\Psi_{\rm{I}}}$ would vanish. However, if the occupied valence subspace of $\ket{\Psi_{F}^{\rm{exact}}}$ has any overlap with the unoccupied subspace of $\ket{\Psi_{\rm{I}}}$, the higher order terms will inevitably appear in Eq.~\ref{FSSD_op}. On the other hand, from Eq.~\ref{TDDFT_FinalSTate}, it is evident that only the first order terms are present in the expansion of $\ket{\Psi_F^{\rm{Resp}}}$. Consequently, in the adiabatic approximation, in which the eigenvectors of $H^{2\rm{p}}$ must be normalized (this is not true for frequency-dependent kernels)~\cite{doi:10.1142/9789812830586_0005}, the set of coefficients $\{\alpha^F\}$ must differ from $\{\gamma^F\}$. This leads to the crucial inference that if there is appreciable difference between the valence occupied subspaces (and consequently, between the valence electronic densities and polarizations) of the initial and the final state, then $M^F_{\rm{Resp}}$, calculated with an adiabatic kernel, will also differ appreciably from $M^F_{\rm{exact}}$. 

The crucial role played by the change in valence electronic polarization is reflected in Fig.~S6,~S7 which show that, for ionizations leading to large change in the valence dipole-moment, the LR-TDDFT spectra differ significantly from the MBXES counterparts, establishing the aforementioned change as a metric for the applicability of LR-TDDFT.  

Note that this inadequacy, which stems from the adiabatic approximation, will be present in any xc-functional and can be mitigated either by employing a sophisticated frequency-dependent kernel, or by including quadratic or higher-order response - both of which are expensive options. Similar limitations of the adiabatic kernel have been noted~\cite{maitra2021double,ELLIOTT2011110} for double-excitations and charge-transfer excitations of valence electrons, in which the localization of electron and hole in different spatial regions gives rise to a large change in density and polarization. 

From Fig~\ref{PhenolSpectra}, we note that, in comparison with the LR-TDDFT spectrum, the single-determinant counterparts are in much better agreement with the experimental spectrum presented in panel (d). In general, the agreement between the eDFT results and the experimental spectra for various molecules indicates that most of the excited states can be  well-approximated by single SDs composed of KS orbitals corresponding to the appropriate SCFs.

\begin{figure}
\centering
\includegraphics[width=0.5\textwidth]{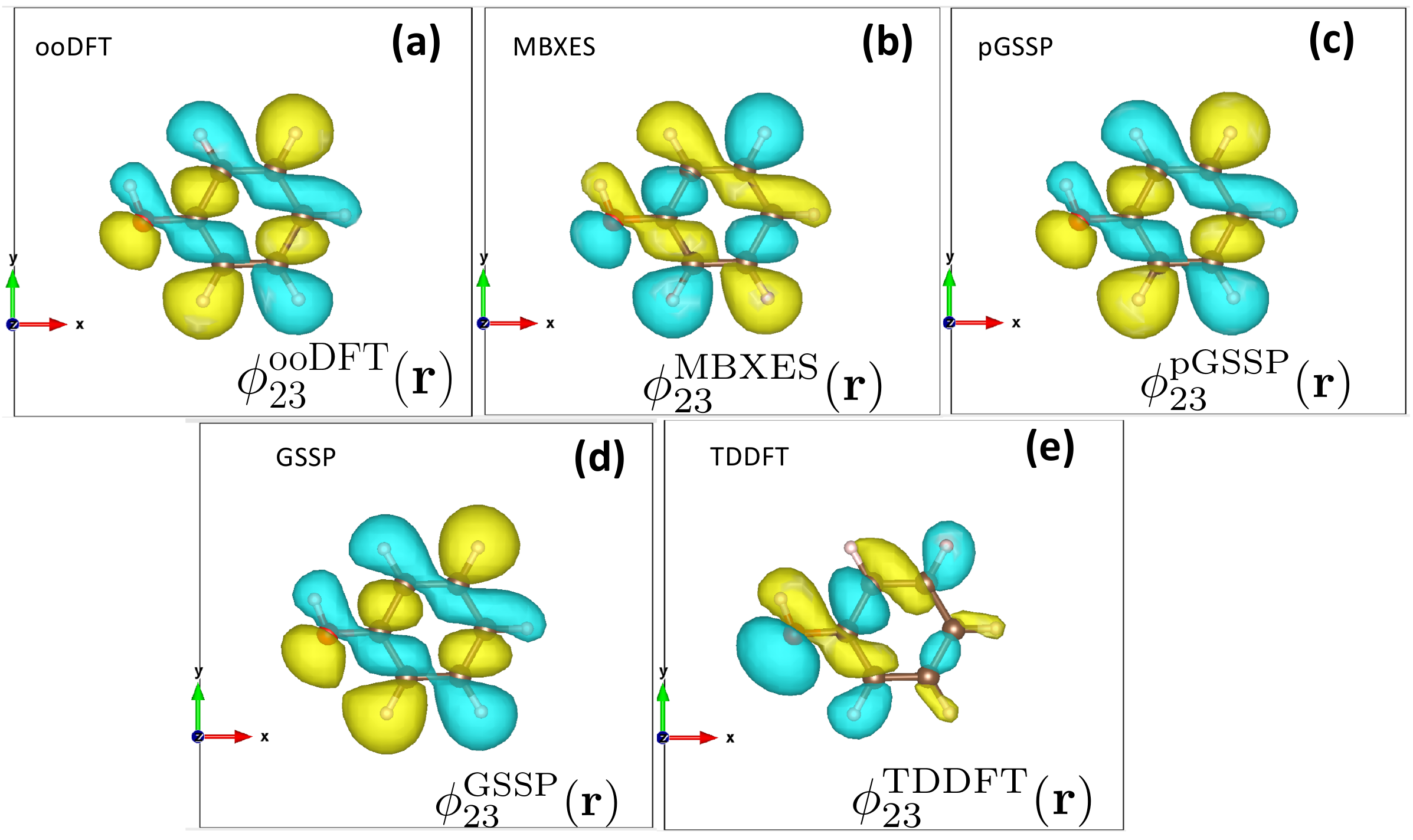}
\caption{Plot of auxiliary orbitals associated with 23rd de-excitation in the (a)ooDFT, (b)MBXES, (c)pGS, (d)GS and (e)LR-TDDFT framework.}\label{OrbitalPlot}
\end{figure}
 
The similarity between the ooDFT  and the MBXES spectra (which, in turn, is in appreciable agreement with the pGS spectrum) in Fig.~\ref{PhenolSpectra} indicates the fact that the removal of a valence electron typically has a relatively low impact on the valence quasiparticle (QP) orbitals and therefore, can often be safely ignored in practical calculations of the oscillator strength. This trend is also observed in the spectra of acetone (Fig.~S2). Conversely, the relative dissimilarity between the MBXES/pGS and the GS spectrum, which is even more conspicuous in the spectra of acetone, points to the crucial fact that, due to the localized nature of a core orbital, addition/removal of a core electron alters the self-consistent field and the valence QP orbitals significantly. It must be noted that, though the simulated spectra are, in principle, dependent on the xc-functional, our general conclusions regarding the different simulation-methods will remain unchanged (see Fig.~S2,~S3,~S4).

As a representative example for the O \textit{K}-edge XES of phenol, we examine the 23rd de-excitation which, as shown by a broken arrow in Fig.~\ref{PhenolSpectra}, exhibits comparable oscillator strengths in the GS, pGS, MBXES and ooDFT treatment. This can be attributed to the similarity (within a global phase-factor of $\pm 1$) among the orbitals $\phi_{23}(\mathbf{r})$ , $\phi^{\rm{pGS}}_{23}(\mathbf{r})$ , $\phi^{\rm{MBXES}}_{23}(\mathbf{r})$ and $\phi^{\rm{ooDFT}}_{23}(\mathbf{r})$ associated with the equivalent single-particle transitions, as shown in Fig.~\ref{OrbitalPlot}(a)-(d). Conversely, the significant difference in $\phi^{\rm{TDDFT}}_{23}(\mathbf{r})$  (panel (e)) is responsible for the appreciable deviation in the TDDFT oscillator strength. A similar trend can be observed in Fig.~S5 which displays, as a function of $j$, the coefficients of $\braket{\phi_c|\hat{o}|\tilde{\phi}_j}$  in the linear expansion of the dipole matrix element for the 23rd de-excitation in the ooDFT ($\Xi^{(23)}_{j,23}$ from  Eq.~\ref{ESD_mat}) , MBXES ($\Xi_{j,23}$ from Eq.~\ref{MBXES_mat}), pGS ($\xi_{j,23}$ from Eq.~\ref{pGS_Mat}),   and LR-TDDFT ($\gamma_j^{(23)}$ from Eq.~\ref{TDDFT_DipoleMat}) treatment, respectively.

Note that a particular advantage of the MBXES method is that unlike the response-based methods, this can be straightforwardly applied in plane-wave pseudopotential calculations for solids. In  this  case,  the  initial-state  calculation  must  be performed on a system where the core-excited atom is represented by a modified pseudopotential. It is expected that, for many systems of interest, among all of the above-mentioned methods, MBXES will strike the ideal balance between accuracy and feasibility, especially for large systems which can render ooDFT formidably expensive. 

Before concluding, we would like to emphasize that, as mentioned earlier, the applicability of the eDFT and the response based treatment is not limited to de-excitations involving the core level. They can, in fact be used to simulate optical de-excitations within the valence subspace as discussed in the SI (see Fig.~S1 for a sample simulated spectrum)

In summary, with systematically tuned approximations, we propose certain theoretical approaches for modeling XES and present an in-depth comparison with existing tools for studying electronic transitions by highlighting the physical approximations contained within each approach. XES can be modeled accurately by obtaining each final state separately using ooDFT. However, owing to the relatively delocalized nature of a valence hole, the much cheaper MBXES method, which requires only a single SCF calculation for all final states, is often adequate for obtaining the relevant oscillator strengths with high accuracy. Furthermore, the pGS approximation, which replaces the overlap of SDs with that of the relevant KS orbitals is reasonably accurate. However, completely neglecting the effects of the core-hole (using the GS transition amplitudes or the pDOS) may lead to deviations. Finally we demonstrate that, unless the core-ionization preserves the valence density appreciably (reflected by a small change in the valence dipole moment), adiabatic linear-response approaches can introduce significant errors and should be used with caution when simulating de-excitations to the core level.

\bibliographystyle{apsrev4-2}
\bibliography{biblio.bib}

\begin{thebibliography}{55}%
\makeatletter
\providecommand \@ifxundefined [1]{%
 \@ifx{#1\undefined}
}%
\providecommand \@ifnum [1]{%
 \ifnum #1\expandafter \@firstoftwo
 \else \expandafter \@secondoftwo
 \fi
}%
\providecommand \@ifx [1]{%
 \ifx #1\expandafter \@firstoftwo
 \else \expandafter \@secondoftwo
 \fi
}%
\providecommand \natexlab [1]{#1}%
\providecommand \enquote  [1]{``#1''}%
\providecommand \bibnamefont  [1]{#1}%
\providecommand \bibfnamefont [1]{#1}%
\providecommand \citenamefont [1]{#1}%
\providecommand \href@noop [0]{\@secondoftwo}%
\providecommand \href [0]{\begingroup \@sanitize@url \@href}%
\providecommand \@href[1]{\@@startlink{#1}\@@href}%
\providecommand \@@href[1]{\endgroup#1\@@endlink}%
\providecommand \@sanitize@url [0]{\catcode `\\12\catcode `\$12\catcode
  `\&12\catcode `\#12\catcode `\^12\catcode `\_12\catcode `\%12\relax}%
\providecommand \@@startlink[1]{}%
\providecommand \@@endlink[0]{}%
\providecommand \url  [0]{\begingroup\@sanitize@url \@url }%
\providecommand \@url [1]{\endgroup\@href {#1}{\urlprefix }}%
\providecommand \urlprefix  [0]{URL }%
\providecommand \Eprint [0]{\href }%
\providecommand \doibase [0]{https://doi.org/}%
\providecommand \selectlanguage [0]{\@gobble}%
\providecommand \bibinfo  [0]{\@secondoftwo}%
\providecommand \bibfield  [0]{\@secondoftwo}%
\providecommand \translation [1]{[#1]}%
\providecommand \BibitemOpen [0]{}%
\providecommand \bibitemStop [0]{}%
\providecommand \bibitemNoStop [0]{.\EOS\space}%
\providecommand \EOS [0]{\spacefactor3000\relax}%
\providecommand \BibitemShut  [1]{\csname bibitem#1\endcsname}%
\let\auto@bib@innerbib\@empty
\bibitem [{\citenamefont {Hohenberg}\ and\ \citenamefont
  {Kohn}(1964)}]{PhysRev.136.B864}%
  \BibitemOpen
  \bibfield  {author} {\bibinfo {author} {\bibfnamefont {P.}~\bibnamefont
  {Hohenberg}}\ and\ \bibinfo {author} {\bibfnamefont {W.}~\bibnamefont
  {Kohn}},\ }\href {https://doi.org/10.1103/PhysRev.136.B864} {\bibfield
  {journal} {\bibinfo  {journal} {Phys. Rev.}\ }\textbf {\bibinfo {volume}
  {136}},\ \bibinfo {pages} {B864} (\bibinfo {year} {1964})}\BibitemShut
  {NoStop}%
\bibitem [{\citenamefont {Kohn}\ and\ \citenamefont
  {Sham}(1965)}]{PhysRev.140.A1133}%
  \BibitemOpen
  \bibfield  {author} {\bibinfo {author} {\bibfnamefont {W.}~\bibnamefont
  {Kohn}}\ and\ \bibinfo {author} {\bibfnamefont {L.~J.}\ \bibnamefont
  {Sham}},\ }\href {https://doi.org/10.1103/PhysRev.140.A1133} {\bibfield
  {journal} {\bibinfo  {journal} {Phys. Rev.}\ }\textbf {\bibinfo {volume}
  {140}},\ \bibinfo {pages} {A1133} (\bibinfo {year} {1965})}\BibitemShut
  {NoStop}%
\bibitem [{\citenamefont {Becke}(2014)}]{doi:10.1063/1.4869598}%
  \BibitemOpen
  \bibfield  {author} {\bibinfo {author} {\bibfnamefont {A.~D.}\ \bibnamefont
  {Becke}},\ }\href {https://doi.org/10.1063/1.4869598} {\bibfield  {journal}
  {\bibinfo  {journal} {J. Chem. Phys.}\ }\textbf {\bibinfo {volume} {140}},\
  \bibinfo {pages} {18A301} (\bibinfo {year} {2014})}\BibitemShut {NoStop}%
\bibitem [{\citenamefont {Jones}(2015)}]{RevModPhys.87.897}%
  \BibitemOpen
  \bibfield  {author} {\bibinfo {author} {\bibfnamefont {R.~O.}\ \bibnamefont
  {Jones}},\ }\href {https://doi.org/10.1103/RevModPhys.87.897} {\bibfield
  {journal} {\bibinfo  {journal} {Rev. Mod. Phys.}\ }\textbf {\bibinfo {volume}
  {87}},\ \bibinfo {pages} {897} (\bibinfo {year} {2015})}\BibitemShut
  {NoStop}%
\bibitem [{\citenamefont {Tozer}\ and\ \citenamefont
  {Peach}(2014)}]{C4CP90074J}%
  \BibitemOpen
  \bibfield  {author} {\bibinfo {author} {\bibfnamefont {D.~J.}\ \bibnamefont
  {Tozer}}\ and\ \bibinfo {author} {\bibfnamefont {M.~J.~G.}\ \bibnamefont
  {Peach}},\ }\href {https://doi.org/10.1039/C4CP90074J} {\bibfield  {journal}
  {\bibinfo  {journal} {Phys. Chem. Chem. Phys.}\ }\textbf {\bibinfo {volume}
  {16}},\ \bibinfo {pages} {14333} (\bibinfo {year} {2014})}\BibitemShut
  {NoStop}%
\bibitem [{\citenamefont {van Mourik}\ \emph {et~al.}(2014)\citenamefont {van
  Mourik}, \citenamefont {Bühl},\ and\ \citenamefont
  {Gaigeot}}]{doi:10.1098/rsta.2012.0488}%
  \BibitemOpen
  \bibfield  {author} {\bibinfo {author} {\bibfnamefont {T.}~\bibnamefont {van
  Mourik}}, \bibinfo {author} {\bibfnamefont {M.}~\bibnamefont {Bühl}},\ and\
  \bibinfo {author} {\bibfnamefont {M.-P.}\ \bibnamefont {Gaigeot}},\ }\href
  {https://doi.org/10.1098/rsta.2012.0488} {\bibfield  {journal} {\bibinfo
  {journal} {Philos. Trans. R. Soc. A}\ }\textbf {\bibinfo {volume} {372}},\
  \bibinfo {pages} {20120488} (\bibinfo {year} {2014})}\BibitemShut {NoStop}%
\bibitem [{\citenamefont {Besley}()}]{besleys}%
  \BibitemOpen
  \bibfield  {author} {\bibinfo {author} {\bibfnamefont {N.~A.}\ \bibnamefont
  {Besley}},\ }\href {https://doi.org/https://doi.org/10.1002/wcms.1527}
  {\bibfield  {journal} {\bibinfo  {journal} {Wiley Interdiscip. Rev. Comput.
  Mol. Sci}\ }\textbf {\bibinfo {volume} {n/a}},\ \bibinfo {pages}
  {e1527}}\BibitemShut {NoStop}%
\bibitem [{\citenamefont {Choudhary}\ \emph {et~al.}(2021)\citenamefont
  {Choudhary}, \citenamefont {Garrity}, \citenamefont {Camp}, \citenamefont
  {Kalinin}, \citenamefont {Vasudevan}, \citenamefont {Ziatdinov},\ and\
  \citenamefont {Tavazza}}]{Choudhary2021}%
  \BibitemOpen
  \bibfield  {author} {\bibinfo {author} {\bibfnamefont {K.}~\bibnamefont
  {Choudhary}}, \bibinfo {author} {\bibfnamefont {K.~F.}\ \bibnamefont
  {Garrity}}, \bibinfo {author} {\bibfnamefont {C.}~\bibnamefont {Camp}},
  \bibinfo {author} {\bibfnamefont {S.~V.}\ \bibnamefont {Kalinin}}, \bibinfo
  {author} {\bibfnamefont {R.}~\bibnamefont {Vasudevan}}, \bibinfo {author}
  {\bibfnamefont {M.}~\bibnamefont {Ziatdinov}},\ and\ \bibinfo {author}
  {\bibfnamefont {F.}~\bibnamefont {Tavazza}},\ }\href
  {https://doi.org/10.1038/s41597-021-00824-y} {\bibfield  {journal} {\bibinfo
  {journal} {Sci. Data}\ }\textbf {\bibinfo {volume} {8}},\ \bibinfo {pages}
  {57} (\bibinfo {year} {2021})}\BibitemShut {NoStop}%
\bibitem [{\citenamefont {Jain}\ \emph {et~al.}(2013)\citenamefont {Jain},
  \citenamefont {Ong}, \citenamefont {Hautier}, \citenamefont {Chen},
  \citenamefont {Richards}, \citenamefont {Dacek}, \citenamefont {Cholia},
  \citenamefont {Gunter}, \citenamefont {Skinner}, \citenamefont {Ceder},\ and\
  \citenamefont {Persson}}]{doi:10.1063/1.4812323}%
  \BibitemOpen
  \bibfield  {author} {\bibinfo {author} {\bibfnamefont {A.}~\bibnamefont
  {Jain}}, \bibinfo {author} {\bibfnamefont {S.~P.}\ \bibnamefont {Ong}},
  \bibinfo {author} {\bibfnamefont {G.}~\bibnamefont {Hautier}}, \bibinfo
  {author} {\bibfnamefont {W.}~\bibnamefont {Chen}}, \bibinfo {author}
  {\bibfnamefont {W.~D.}\ \bibnamefont {Richards}}, \bibinfo {author}
  {\bibfnamefont {S.}~\bibnamefont {Dacek}}, \bibinfo {author} {\bibfnamefont
  {S.}~\bibnamefont {Cholia}}, \bibinfo {author} {\bibfnamefont
  {D.}~\bibnamefont {Gunter}}, \bibinfo {author} {\bibfnamefont
  {D.}~\bibnamefont {Skinner}}, \bibinfo {author} {\bibfnamefont
  {G.}~\bibnamefont {Ceder}},\ and\ \bibinfo {author} {\bibfnamefont {K.~A.}\
  \bibnamefont {Persson}},\ }\href {https://doi.org/10.1063/1.4812323}
  {\bibfield  {journal} {\bibinfo  {journal} {APL Mater.}\ }\textbf {\bibinfo
  {volume} {1}},\ \bibinfo {pages} {011002} (\bibinfo {year}
  {2013})}\BibitemShut {NoStop}%
\bibitem [{\citenamefont {Curtarolo}\ \emph {et~al.}(2013)\citenamefont
  {Curtarolo}, \citenamefont {Hart}, \citenamefont {Nardelli}, \citenamefont
  {Mingo}, \citenamefont {Sanvito},\ and\ \citenamefont
  {Levy}}]{Curtarolo2013}%
  \BibitemOpen
  \bibfield  {author} {\bibinfo {author} {\bibfnamefont {S.}~\bibnamefont
  {Curtarolo}}, \bibinfo {author} {\bibfnamefont {G.~L.~W.}\ \bibnamefont
  {Hart}}, \bibinfo {author} {\bibfnamefont {M.~B.}\ \bibnamefont {Nardelli}},
  \bibinfo {author} {\bibfnamefont {N.}~\bibnamefont {Mingo}}, \bibinfo
  {author} {\bibfnamefont {S.}~\bibnamefont {Sanvito}},\ and\ \bibinfo {author}
  {\bibfnamefont {O.}~\bibnamefont {Levy}},\ }\href
  {https://doi.org/10.1038/nmat3568} {\bibfield  {journal} {\bibinfo  {journal}
  {Nat. Mater.}\ }\textbf {\bibinfo {volume} {12}},\ \bibinfo {pages} {191}
  (\bibinfo {year} {2013})}\BibitemShut {NoStop}%
\bibitem [{\citenamefont {Hait}\ and\ \citenamefont
  {Head-Gordon}(2021)}]{Hait2021}%
  \BibitemOpen
  \bibfield  {author} {\bibinfo {author} {\bibfnamefont {D.}~\bibnamefont
  {Hait}}\ and\ \bibinfo {author} {\bibfnamefont {M.}~\bibnamefont
  {Head-Gordon}},\ }\href {https://doi.org/10.1021/acs.jpclett.1c00744}
  {\bibfield  {journal} {\bibinfo  {journal} {J. Phys. Chem. Lett.}\ }\textbf
  {\bibinfo {volume} {12}},\ \bibinfo {pages} {4517} (\bibinfo {year}
  {2021})}\BibitemShut {NoStop}%
\bibitem [{\citenamefont {Cullen}\ \emph {et~al.}(2011)\citenamefont {Cullen},
  \citenamefont {Krykunov},\ and\ \citenamefont {Ziegler}}]{CULLEN201111}%
  \BibitemOpen
  \bibfield  {author} {\bibinfo {author} {\bibfnamefont {J.}~\bibnamefont
  {Cullen}}, \bibinfo {author} {\bibfnamefont {M.}~\bibnamefont {Krykunov}},\
  and\ \bibinfo {author} {\bibfnamefont {T.}~\bibnamefont {Ziegler}},\ }\href
  {https://doi.org/https://doi.org/10.1016/j.chemphys.2011.05.021} {\bibfield
  {journal} {\bibinfo  {journal} {Chem. Phys.}\ }\textbf {\bibinfo {volume}
  {391}},\ \bibinfo {pages} {11} (\bibinfo {year} {2011})}\BibitemShut
  {NoStop}%
\bibitem [{\citenamefont {Evangelista}\ \emph {et~al.}(2013)\citenamefont
  {Evangelista}, \citenamefont {Shushkov},\ and\ \citenamefont
  {Tully}}]{doi:10.1021/jp401323d}%
  \BibitemOpen
  \bibfield  {author} {\bibinfo {author} {\bibfnamefont {F.~A.}\ \bibnamefont
  {Evangelista}}, \bibinfo {author} {\bibfnamefont {P.}~\bibnamefont
  {Shushkov}},\ and\ \bibinfo {author} {\bibfnamefont {J.~C.}\ \bibnamefont
  {Tully}},\ }\href@noop {} {\bibfield  {journal} {\bibinfo  {journal} {J.
  Phys. Chem. A .}\ }\textbf {\bibinfo {volume} {117}},\ \bibinfo {pages}
  {7378} (\bibinfo {year} {2013})}\BibitemShut {NoStop}%
\bibitem [{\citenamefont {Levi}\ \emph {et~al.}(2020)\citenamefont {Levi},
  \citenamefont {Ivanov},\ and\ \citenamefont
  {Jónsson}}]{doi:10.1021/acs.jctc.0c00597}%
  \BibitemOpen
  \bibfield  {author} {\bibinfo {author} {\bibfnamefont {G.}~\bibnamefont
  {Levi}}, \bibinfo {author} {\bibfnamefont {A.~V.}\ \bibnamefont {Ivanov}},\
  and\ \bibinfo {author} {\bibfnamefont {H.}~\bibnamefont {Jónsson}},\
  }\href@noop {} {\bibfield  {journal} {\bibinfo  {journal} {J. Chem. Theory
  Comput.}\ }\textbf {\bibinfo {volume} {16}},\ \bibinfo {pages} {6968}
  (\bibinfo {year} {2020})}\BibitemShut {NoStop}%
\bibitem [{\citenamefont {Ramos}\ and\ \citenamefont
  {Pavanello}(2018)}]{doi:10.1063/1.5018615}%
  \BibitemOpen
  \bibfield  {author} {\bibinfo {author} {\bibfnamefont {P.}~\bibnamefont
  {Ramos}}\ and\ \bibinfo {author} {\bibfnamefont {M.}~\bibnamefont
  {Pavanello}},\ }\href@noop {} {\bibfield  {journal} {\bibinfo  {journal} {J.
  Chem. Phys.}\ }\textbf {\bibinfo {volume} {148}},\ \bibinfo {pages} {144103}
  (\bibinfo {year} {2018})}\BibitemShut {NoStop}%
\bibitem [{\citenamefont {Cheng}\ \emph {et~al.}(2008)\citenamefont {Cheng},
  \citenamefont {Wu},\ and\ \citenamefont
  {Van~Voorhis}}]{doi:10.1063/1.2977989}%
  \BibitemOpen
  \bibfield  {author} {\bibinfo {author} {\bibfnamefont {C.-L.}\ \bibnamefont
  {Cheng}}, \bibinfo {author} {\bibfnamefont {Q.}~\bibnamefont {Wu}},\ and\
  \bibinfo {author} {\bibfnamefont {T.}~\bibnamefont {Van~Voorhis}},\
  }\href@noop {} {\bibfield  {journal} {\bibinfo  {journal} {J. Chem. Phys.}\
  }\textbf {\bibinfo {volume} {129}},\ \bibinfo {pages} {124112} (\bibinfo
  {year} {2008})}\BibitemShut {NoStop}%
\bibitem [{\citenamefont {Roychoudhury}\ \emph {et~al.}(2020)\citenamefont
  {Roychoudhury}, \citenamefont {Sanvito},\ and\ \citenamefont
  {O'Regan}}]{Roychoudhury2020}%
  \BibitemOpen
  \bibfield  {author} {\bibinfo {author} {\bibfnamefont {S.}~\bibnamefont
  {Roychoudhury}}, \bibinfo {author} {\bibfnamefont {S.}~\bibnamefont
  {Sanvito}},\ and\ \bibinfo {author} {\bibfnamefont {D.~D.}\ \bibnamefont
  {O'Regan}},\ }\href {https://doi.org/10.1038/s41598-020-65209-4} {\bibfield
  {journal} {\bibinfo  {journal} {Sci. Rep.}\ }\textbf {\bibinfo {volume}
  {10}},\ \bibinfo {pages} {8947} (\bibinfo {year} {2020})}\BibitemShut
  {NoStop}%
\bibitem [{\citenamefont {Petersilka}\ \emph {et~al.}(1996)\citenamefont
  {Petersilka}, \citenamefont {Gossmann},\ and\ \citenamefont
  {Gross}}]{PhysRevLett.76.1212}%
  \BibitemOpen
  \bibfield  {author} {\bibinfo {author} {\bibfnamefont {M.}~\bibnamefont
  {Petersilka}}, \bibinfo {author} {\bibfnamefont {U.~J.}\ \bibnamefont
  {Gossmann}},\ and\ \bibinfo {author} {\bibfnamefont {E.~K.~U.}\ \bibnamefont
  {Gross}},\ }\href {https://doi.org/10.1103/PhysRevLett.76.1212} {\bibfield
  {journal} {\bibinfo  {journal} {Phys. Rev. Lett.}\ }\textbf {\bibinfo
  {volume} {76}},\ \bibinfo {pages} {1212} (\bibinfo {year}
  {1996})}\BibitemShut {NoStop}%
\bibitem [{\citenamefont {Casida}()}]{doi:10.1142/9789812830586_0005}%
  \BibitemOpen
  \bibfield  {author} {\bibinfo {author} {\bibfnamefont {M.~E.}\ \bibnamefont
  {Casida}},\ }\bibinfo {title} {Time-dependent density functional response
  theory for molecules},\ in\ \href
  {https://doi.org/10.1142/9789812830586_0005} {\emph {\bibinfo {booktitle}
  {Recent Advances in Density Functional Methods}}},\ pp.\ \bibinfo {pages}
  {155--192}\BibitemShut {NoStop}%
\bibitem [{\citenamefont {Shirley}(1998)}]{PhysRevLett.80.794}%
  \BibitemOpen
  \bibfield  {author} {\bibinfo {author} {\bibfnamefont {E.~L.}\ \bibnamefont
  {Shirley}},\ }\href {https://doi.org/10.1103/PhysRevLett.80.794} {\bibfield
  {journal} {\bibinfo  {journal} {Phys. Rev. Lett.}\ }\textbf {\bibinfo
  {volume} {80}},\ \bibinfo {pages} {794} (\bibinfo {year} {1998})}\BibitemShut
  {NoStop}%
\bibitem [{\citenamefont {Rohlfing}\ and\ \citenamefont
  {Louie}(2000)}]{PhysRevB.62.4927}%
  \BibitemOpen
  \bibfield  {author} {\bibinfo {author} {\bibfnamefont {M.}~\bibnamefont
  {Rohlfing}}\ and\ \bibinfo {author} {\bibfnamefont {S.~G.}\ \bibnamefont
  {Louie}},\ }\href {https://doi.org/10.1103/PhysRevB.62.4927} {\bibfield
  {journal} {\bibinfo  {journal} {Phys. Rev. B}\ }\textbf {\bibinfo {volume}
  {62}},\ \bibinfo {pages} {4927} (\bibinfo {year} {2000})}\BibitemShut
  {NoStop}%
\bibitem [{\citenamefont {Onida}\ \emph {et~al.}(2002)\citenamefont {Onida},
  \citenamefont {Reining},\ and\ \citenamefont {Rubio}}]{RevModPhys.74.601}%
  \BibitemOpen
  \bibfield  {author} {\bibinfo {author} {\bibfnamefont {G.}~\bibnamefont
  {Onida}}, \bibinfo {author} {\bibfnamefont {L.}~\bibnamefont {Reining}},\
  and\ \bibinfo {author} {\bibfnamefont {A.}~\bibnamefont {Rubio}},\ }\href
  {https://doi.org/10.1103/RevModPhys.74.601} {\bibfield  {journal} {\bibinfo
  {journal} {Rev. Mod. Phys.}\ }\textbf {\bibinfo {volume} {74}},\ \bibinfo
  {pages} {601} (\bibinfo {year} {2002})}\BibitemShut {NoStop}%
\bibitem [{\citenamefont {Kotani}(1996)}]{KOTANI19967}%
  \BibitemOpen
  \bibfield  {author} {\bibinfo {author} {\bibfnamefont {A.}~\bibnamefont
  {Kotani}},\ }\href
  {https://doi.org/https://doi.org/10.1016/S0368-2048(96)80017-X} {\bibfield
  {journal} {\bibinfo  {journal} {J. Electron Spectrosc. Relat. Phenom.}\
  }\textbf {\bibinfo {volume} {78}},\ \bibinfo {pages} {7} (\bibinfo {year}
  {1996})}\BibitemShut {NoStop}%
\bibitem [{\citenamefont {de~Groot}(2001)}]{doi:10.1021/cr9900681}%
  \BibitemOpen
  \bibfield  {author} {\bibinfo {author} {\bibfnamefont {F.}~\bibnamefont
  {de~Groot}},\ }\href@noop {} {\bibfield  {journal} {\bibinfo  {journal}
  {Chem. Rev.}\ }\textbf {\bibinfo {volume} {101}},\ \bibinfo {pages} {1779}
  (\bibinfo {year} {2001})}\BibitemShut {NoStop}%
\bibitem [{\citenamefont {Grünert}\ and\ \citenamefont
  {Klementiev}(2020)}]{GrunertKlementiev+2020}%
  \BibitemOpen
  \bibfield  {author} {\bibinfo {author} {\bibfnamefont {W.}~\bibnamefont
  {Grünert}}\ and\ \bibinfo {author} {\bibfnamefont {K.}~\bibnamefont
  {Klementiev}},\ }\href {https://doi.org/doi:10.1515/psr-2017-0181} {\bibfield
   {journal} {\bibinfo  {journal} {Phys. Sci. Rev.}\ }\textbf {\bibinfo
  {volume} {5}},\ \bibinfo {pages} {20170181} (\bibinfo {year}
  {2020})}\BibitemShut {NoStop}%
\bibitem [{\citenamefont {Vinson}\ \emph {et~al.}(2011)\citenamefont {Vinson},
  \citenamefont {Rehr}, \citenamefont {Kas},\ and\ \citenamefont
  {Shirley}}]{PhysRevB.83.115106}%
  \BibitemOpen
  \bibfield  {author} {\bibinfo {author} {\bibfnamefont {J.}~\bibnamefont
  {Vinson}}, \bibinfo {author} {\bibfnamefont {J.~J.}\ \bibnamefont {Rehr}},
  \bibinfo {author} {\bibfnamefont {J.~J.}\ \bibnamefont {Kas}},\ and\ \bibinfo
  {author} {\bibfnamefont {E.~L.}\ \bibnamefont {Shirley}},\ }\href
  {https://doi.org/10.1103/PhysRevB.83.115106} {\bibfield  {journal} {\bibinfo
  {journal} {Phys. Rev. B}\ }\textbf {\bibinfo {volume} {83}},\ \bibinfo
  {pages} {115106} (\bibinfo {year} {2011})}\BibitemShut {NoStop}%
\bibitem [{\citenamefont {von Barth}\ and\ \citenamefont
  {Grossmann}(1982)}]{PhysRevB.25.5150}%
  \BibitemOpen
  \bibfield  {author} {\bibinfo {author} {\bibfnamefont {U.}~\bibnamefont {von
  Barth}}\ and\ \bibinfo {author} {\bibfnamefont {G.}~\bibnamefont
  {Grossmann}},\ }\href {https://doi.org/10.1103/PhysRevB.25.5150} {\bibfield
  {journal} {\bibinfo  {journal} {Phys. Rev. B}\ }\textbf {\bibinfo {volume}
  {25}},\ \bibinfo {pages} {5150} (\bibinfo {year} {1982})}\BibitemShut
  {NoStop}%
\bibitem [{\citenamefont {Rehr}\ \emph {et~al.}(2009)\citenamefont {Rehr},
  \citenamefont {Kas}, \citenamefont {Prange}, \citenamefont {Sorini},
  \citenamefont {Takimoto},\ and\ \citenamefont {Vila}}]{REHR2009548}%
  \BibitemOpen
  \bibfield  {author} {\bibinfo {author} {\bibfnamefont {J.~J.}\ \bibnamefont
  {Rehr}}, \bibinfo {author} {\bibfnamefont {J.~J.}\ \bibnamefont {Kas}},
  \bibinfo {author} {\bibfnamefont {M.~P.}\ \bibnamefont {Prange}}, \bibinfo
  {author} {\bibfnamefont {A.~P.}\ \bibnamefont {Sorini}}, \bibinfo {author}
  {\bibfnamefont {Y.}~\bibnamefont {Takimoto}},\ and\ \bibinfo {author}
  {\bibfnamefont {F.}~\bibnamefont {Vila}},\ }\href
  {https://doi.org/https://doi.org/10.1016/j.crhy.2008.08.004} {\bibfield
  {journal} {\bibinfo  {journal} {C. R. Phys.}\ }\textbf {\bibinfo {volume}
  {10}},\ \bibinfo {pages} {548} (\bibinfo {year} {2009})},\ \bibinfo {note}
  {theoretical spectroscopy}\BibitemShut {NoStop}%
\bibitem [{\citenamefont {Besley}(2020)}]{doi:10.1021/acs.accounts.0c00171}%
  \BibitemOpen
  \bibfield  {author} {\bibinfo {author} {\bibfnamefont {N.~A.}\ \bibnamefont
  {Besley}},\ }\href@noop {} {\bibfield  {journal} {\bibinfo  {journal} {Acc.
  Chem. Res}\ }\textbf {\bibinfo {volume} {53}},\ \bibinfo {pages} {1306}
  (\bibinfo {year} {2020})}\BibitemShut {NoStop}%
\bibitem [{\citenamefont {Fransson}\ and\ \citenamefont
  {Dreuw}(2019)}]{doi:10.1021/acs.jctc.8b01046}%
  \BibitemOpen
  \bibfield  {author} {\bibinfo {author} {\bibfnamefont {T.}~\bibnamefont
  {Fransson}}\ and\ \bibinfo {author} {\bibfnamefont {A.}~\bibnamefont
  {Dreuw}},\ }\href@noop {} {\bibfield  {journal} {\bibinfo  {journal} {J.
  Chem. Theory Comput.}\ }\textbf {\bibinfo {volume} {15}},\ \bibinfo {pages}
  {546} (\bibinfo {year} {2019})}\BibitemShut {NoStop}%
\bibitem [{\citenamefont {Vorwerk}\ \emph {et~al.}(2020)\citenamefont
  {Vorwerk}, \citenamefont {Sottile},\ and\ \citenamefont
  {Draxl}}]{PhysRevResearch.2.042003}%
  \BibitemOpen
  \bibfield  {author} {\bibinfo {author} {\bibfnamefont {C.}~\bibnamefont
  {Vorwerk}}, \bibinfo {author} {\bibfnamefont {F.}~\bibnamefont {Sottile}},\
  and\ \bibinfo {author} {\bibfnamefont {C.}~\bibnamefont {Draxl}},\ }\href
  {https://doi.org/10.1103/PhysRevResearch.2.042003} {\bibfield  {journal}
  {\bibinfo  {journal} {Phys. Rev. Research}\ }\textbf {\bibinfo {volume}
  {2}},\ \bibinfo {pages} {042003} (\bibinfo {year} {2020})}\BibitemShut
  {NoStop}%
\bibitem [{\citenamefont {Liang}\ \emph {et~al.}(2017)\citenamefont {Liang},
  \citenamefont {Vinson}, \citenamefont {Pemmaraju}, \citenamefont {Drisdell},
  \citenamefont {Shirley},\ and\ \citenamefont
  {Prendergast}}]{PhysRevLett.118.096402}%
  \BibitemOpen
  \bibfield  {author} {\bibinfo {author} {\bibfnamefont {Y.}~\bibnamefont
  {Liang}}, \bibinfo {author} {\bibfnamefont {J.}~\bibnamefont {Vinson}},
  \bibinfo {author} {\bibfnamefont {S.}~\bibnamefont {Pemmaraju}}, \bibinfo
  {author} {\bibfnamefont {W.~S.}\ \bibnamefont {Drisdell}}, \bibinfo {author}
  {\bibfnamefont {E.~L.}\ \bibnamefont {Shirley}},\ and\ \bibinfo {author}
  {\bibfnamefont {D.}~\bibnamefont {Prendergast}},\ }\href
  {https://doi.org/10.1103/PhysRevLett.118.096402} {\bibfield  {journal}
  {\bibinfo  {journal} {Phys. Rev. Lett.}\ }\textbf {\bibinfo {volume} {118}},\
  \bibinfo {pages} {096402} (\bibinfo {year} {2017})}\BibitemShut {NoStop}%
\bibitem [{\citenamefont {Vorwerk}\ \emph {et~al.}(2017)\citenamefont
  {Vorwerk}, \citenamefont {Cocchi},\ and\ \citenamefont
  {Draxl}}]{PhysRevB.95.155121}%
  \BibitemOpen
  \bibfield  {author} {\bibinfo {author} {\bibfnamefont {C.}~\bibnamefont
  {Vorwerk}}, \bibinfo {author} {\bibfnamefont {C.}~\bibnamefont {Cocchi}},\
  and\ \bibinfo {author} {\bibfnamefont {C.}~\bibnamefont {Draxl}},\ }\href
  {https://doi.org/10.1103/PhysRevB.95.155121} {\bibfield  {journal} {\bibinfo
  {journal} {Phys. Rev. B}\ }\textbf {\bibinfo {volume} {95}},\ \bibinfo
  {pages} {155121} (\bibinfo {year} {2017})}\BibitemShut {NoStop}%
\bibitem [{\citenamefont {Oosterbaan}\ \emph {et~al.}(2020)\citenamefont
  {Oosterbaan}, \citenamefont {White}, \citenamefont {Hait},\ and\
  \citenamefont {Head-Gordon}}]{C9CP06592J}%
  \BibitemOpen
  \bibfield  {author} {\bibinfo {author} {\bibfnamefont {K.~J.}\ \bibnamefont
  {Oosterbaan}}, \bibinfo {author} {\bibfnamefont {A.~F.}\ \bibnamefont
  {White}}, \bibinfo {author} {\bibfnamefont {D.}~\bibnamefont {Hait}},\ and\
  \bibinfo {author} {\bibfnamefont {M.}~\bibnamefont {Head-Gordon}},\ }\href
  {https://doi.org/10.1039/C9CP06592J} {\bibfield  {journal} {\bibinfo
  {journal} {Phys. Chem. Chem. Phys.}\ }\textbf {\bibinfo {volume} {22}},\
  \bibinfo {pages} {8182} (\bibinfo {year} {2020})}\BibitemShut {NoStop}%
\bibitem [{\citenamefont {Ament}\ \emph {et~al.}(2011)\citenamefont {Ament},
  \citenamefont {van Veenendaal}, \citenamefont {Devereaux}, \citenamefont
  {Hill},\ and\ \citenamefont {van~den Brink}}]{RevModPhys.83.705}%
  \BibitemOpen
  \bibfield  {author} {\bibinfo {author} {\bibfnamefont {L.~J.~P.}\
  \bibnamefont {Ament}}, \bibinfo {author} {\bibfnamefont {M.}~\bibnamefont
  {van Veenendaal}}, \bibinfo {author} {\bibfnamefont {T.~P.}\ \bibnamefont
  {Devereaux}}, \bibinfo {author} {\bibfnamefont {J.~P.}\ \bibnamefont
  {Hill}},\ and\ \bibinfo {author} {\bibfnamefont {J.}~\bibnamefont {van~den
  Brink}},\ }\href {https://doi.org/10.1103/RevModPhys.83.705} {\bibfield
  {journal} {\bibinfo  {journal} {Rev. Mod. Phys.}\ }\textbf {\bibinfo {volume}
  {83}},\ \bibinfo {pages} {705} (\bibinfo {year} {2011})}\BibitemShut
  {NoStop}%
\bibitem [{\citenamefont {de~Groot}\ \emph {et~al.}(1990)\citenamefont
  {de~Groot}, \citenamefont {Fuggle}, \citenamefont {Thole},\ and\
  \citenamefont {Sawatzky}}]{PhysRevB.42.5459}%
  \BibitemOpen
  \bibfield  {author} {\bibinfo {author} {\bibfnamefont {F.~M.~F.}\
  \bibnamefont {de~Groot}}, \bibinfo {author} {\bibfnamefont {J.~C.}\
  \bibnamefont {Fuggle}}, \bibinfo {author} {\bibfnamefont {B.~T.}\
  \bibnamefont {Thole}},\ and\ \bibinfo {author} {\bibfnamefont {G.~A.}\
  \bibnamefont {Sawatzky}},\ }\href {https://doi.org/10.1103/PhysRevB.42.5459}
  {\bibfield  {journal} {\bibinfo  {journal} {Phys. Rev. B}\ }\textbf {\bibinfo
  {volume} {42}},\ \bibinfo {pages} {5459} (\bibinfo {year}
  {1990})}\BibitemShut {NoStop}%
\bibitem [{\citenamefont {Aoki}\ and\ \citenamefont
  {Ohno}(2019)}]{PhysRevB.100.075149}%
  \BibitemOpen
  \bibfield  {author} {\bibinfo {author} {\bibfnamefont {T.}~\bibnamefont
  {Aoki}}\ and\ \bibinfo {author} {\bibfnamefont {K.}~\bibnamefont {Ohno}},\
  }\href {https://doi.org/10.1103/PhysRevB.100.075149} {\bibfield  {journal}
  {\bibinfo  {journal} {Phys. Rev. B}\ }\textbf {\bibinfo {volume} {100}},\
  \bibinfo {pages} {075149} (\bibinfo {year} {2019})}\BibitemShut {NoStop}%
\bibitem [{\citenamefont {Hait}\ and\ \citenamefont
  {Head-Gordon}(2020)}]{doi:10.1021/acs.jpclett.9b03661}%
  \BibitemOpen
  \bibfield  {author} {\bibinfo {author} {\bibfnamefont {D.}~\bibnamefont
  {Hait}}\ and\ \bibinfo {author} {\bibfnamefont {M.}~\bibnamefont
  {Head-Gordon}},\ }\href {https://doi.org/10.1021/acs.jpclett.9b03661}
  {\bibfield  {journal} {\bibinfo  {journal} {J. Phys. Chem. Lett.}\ }\textbf
  {\bibinfo {volume} {11}},\ \bibinfo {pages} {775} (\bibinfo {year}
  {2020})}\BibitemShut {NoStop}%
\bibitem [{\citenamefont {Bergmann}\ and\ \citenamefont
  {Glatzel}(2009)}]{Bergmann2009}%
  \BibitemOpen
  \bibfield  {author} {\bibinfo {author} {\bibfnamefont {U.}~\bibnamefont
  {Bergmann}}\ and\ \bibinfo {author} {\bibfnamefont {P.}~\bibnamefont
  {Glatzel}},\ }\href {https://doi.org/10.1007/s11120-009-9483-6} {\bibfield
  {journal} {\bibinfo  {journal} {Photosynth. Res.}\ }\textbf {\bibinfo
  {volume} {102}},\ \bibinfo {pages} {255} (\bibinfo {year}
  {2009})}\BibitemShut {NoStop}%
\bibitem [{\citenamefont {Fouda}\ and\ \citenamefont
  {Besley}(2020)}]{https://doi.org/10.1002/jcc.26153}%
  \BibitemOpen
  \bibfield  {author} {\bibinfo {author} {\bibfnamefont {A.~A.~E.}\
  \bibnamefont {Fouda}}\ and\ \bibinfo {author} {\bibfnamefont {N.~A.}\
  \bibnamefont {Besley}},\ }\href
  {https://doi.org/https://doi.org/10.1002/jcc.26153} {\bibfield  {journal}
  {\bibinfo  {journal} {J. Comput. Chem.}\ }\textbf {\bibinfo {volume} {41}},\
  \bibinfo {pages} {1081} (\bibinfo {year} {2020})}\BibitemShut {NoStop}%
\bibitem [{\citenamefont {Hanson-Heine}\ \emph {et~al.}(2017)\citenamefont
  {Hanson-Heine}, \citenamefont {George},\ and\ \citenamefont
  {Besley}}]{doi:10.1063/1.4977178}%
  \BibitemOpen
  \bibfield  {author} {\bibinfo {author} {\bibfnamefont {M.~W.~D.}\
  \bibnamefont {Hanson-Heine}}, \bibinfo {author} {\bibfnamefont {M.~W.}\
  \bibnamefont {George}},\ and\ \bibinfo {author} {\bibfnamefont {N.~A.}\
  \bibnamefont {Besley}},\ }\href {https://doi.org/10.1063/1.4977178}
  {\bibfield  {journal} {\bibinfo  {journal} {J. Chem. Phys.}\ }\textbf
  {\bibinfo {volume} {146}},\ \bibinfo {pages} {094106} (\bibinfo {year}
  {2017})}\BibitemShut {NoStop}%
\bibitem [{\citenamefont {Gilbert}\ \emph {et~al.}(2008)\citenamefont
  {Gilbert}, \citenamefont {Besley},\ and\ \citenamefont
  {Gill}}]{doi:10.1021/jp801738f}%
  \BibitemOpen
  \bibfield  {author} {\bibinfo {author} {\bibfnamefont {A.~T.~B.}\
  \bibnamefont {Gilbert}}, \bibinfo {author} {\bibfnamefont {N.~A.}\
  \bibnamefont {Besley}},\ and\ \bibinfo {author} {\bibfnamefont {P.~M.~W.}\
  \bibnamefont {Gill}},\ }\href {https://doi.org/10.1021/jp801738f} {\bibfield
  {journal} {\bibinfo  {journal} {J. Phys. Chem. A .}\ }\textbf {\bibinfo
  {volume} {112}},\ \bibinfo {pages} {13164} (\bibinfo {year}
  {2008})}\BibitemShut {NoStop}%
\bibitem [{\citenamefont {Epifanovsky}\ \emph {et~al.}(2021)\citenamefont
  {Epifanovsky}, \citenamefont {Gilbert}, \citenamefont {Feng}, \citenamefont
  {Lee}, \citenamefont {Mao}, \citenamefont {Mardirossian}, \citenamefont
  {Pokhilko}, \citenamefont {White}, \citenamefont {Coons}, \citenamefont
  {Dempwolff}, \citenamefont {Gan}, \citenamefont {Hait}, \citenamefont {Horn},
  \citenamefont {Jacobson}, \citenamefont {Kaliman},  \emph
  {et~al.}}]{doi:10.1063/5.0055522}%
  \BibitemOpen
  \bibfield  {author} {\bibinfo {author} {\bibfnamefont {E.}~\bibnamefont
  {Epifanovsky}}, \bibinfo {author} {\bibfnamefont {A.~T.~B.}\ \bibnamefont
  {Gilbert}}, \bibinfo {author} {\bibfnamefont {X.}~\bibnamefont {Feng}},
  \bibinfo {author} {\bibfnamefont {J.}~\bibnamefont {Lee}}, \bibinfo {author}
  {\bibfnamefont {Y.}~\bibnamefont {Mao}}, \bibinfo {author} {\bibfnamefont
  {N.}~\bibnamefont {Mardirossian}}, \bibinfo {author} {\bibfnamefont
  {P.}~\bibnamefont {Pokhilko}}, \bibinfo {author} {\bibfnamefont {A.~F.}\
  \bibnamefont {White}}, \bibinfo {author} {\bibfnamefont {M.~P.}\ \bibnamefont
  {Coons}}, \bibinfo {author} {\bibfnamefont {A.~L.}\ \bibnamefont
  {Dempwolff}}, \bibinfo {author} {\bibfnamefont {Z.}~\bibnamefont {Gan}},
  \bibinfo {author} {\bibfnamefont {D.}~\bibnamefont {Hait}}, \bibinfo {author}
  {\bibfnamefont {P.~R.}\ \bibnamefont {Horn}}, \bibinfo {author}
  {\bibfnamefont {L.~D.}\ \bibnamefont {Jacobson}}, \bibinfo {author}
  {\bibfnamefont {I.}~\bibnamefont {Kaliman}}, , \emph {et~al.},\ }\href
  {https://doi.org/10.1063/5.0055522} {\bibfield  {journal} {\bibinfo
  {journal} {J. Chem. Phys.}\ }\textbf {\bibinfo {volume} {155}},\ \bibinfo
  {pages} {084801} (\bibinfo {year} {2021})}\BibitemShut {NoStop}%
\bibitem [{\citenamefont {Mardirossian}\ and\ \citenamefont
  {Head-Gordon}(2016)}]{doi:10.1063/1.4952647}%
  \BibitemOpen
  \bibfield  {author} {\bibinfo {author} {\bibfnamefont {N.}~\bibnamefont
  {Mardirossian}}\ and\ \bibinfo {author} {\bibfnamefont {M.}~\bibnamefont
  {Head-Gordon}},\ }\href {https://doi.org/10.1063/1.4952647} {\bibfield
  {journal} {\bibinfo  {journal} {J. Chem. Phys.}\ }\textbf {\bibinfo {volume}
  {144}},\ \bibinfo {pages} {214110} (\bibinfo {year} {2016})}\BibitemShut
  {NoStop}%
\bibitem [{Note1()}]{Note1}%
  \BibitemOpen
  \bibinfo {note} {In panels (b)-(c), a rigid shift has been applied to the
  spectra so that the highest-frequency point coincides with the energy $\left
  (E_{\protect \rm {I}} - E_{\protect \rm {GS}}^{+1}\right )$, where
  $E_{\protect \rm {GS}}^{+1}$ is the total energy of the valence cationic
  ground state.}\BibitemShut {Stop}%
\bibitem [{\citenamefont {Yumatov}\ \emph {et~al.}(1997)\citenamefont
  {Yumatov}, \citenamefont {Okotrub}, \citenamefont {Furin},\ and\
  \citenamefont {Salakhutdinov}}]{Yumatov1997}%
  \BibitemOpen
  \bibfield  {author} {\bibinfo {author} {\bibfnamefont {V.~D.}\ \bibnamefont
  {Yumatov}}, \bibinfo {author} {\bibfnamefont {A.~V.}\ \bibnamefont
  {Okotrub}}, \bibinfo {author} {\bibfnamefont {G.~G.}\ \bibnamefont {Furin}},\
  and\ \bibinfo {author} {\bibfnamefont {N.~F.}\ \bibnamefont
  {Salakhutdinov}},\ }\href {https://doi.org/10.1007/BF02495254} {\bibfield
  {journal} {\bibinfo  {journal} {Russ. Chem. Bull.}\ }\textbf {\bibinfo
  {volume} {46}},\ \bibinfo {pages} {2074} (\bibinfo {year}
  {1997})}\BibitemShut {NoStop}%
\bibitem [{\citenamefont {Liang}\ and\ \citenamefont
  {Prendergast}(2018)}]{PhysRevB.97.205127}%
  \BibitemOpen
  \bibfield  {author} {\bibinfo {author} {\bibfnamefont {Y.}~\bibnamefont
  {Liang}}\ and\ \bibinfo {author} {\bibfnamefont {D.}~\bibnamefont
  {Prendergast}},\ }\href {https://doi.org/10.1103/PhysRevB.97.205127}
  {\bibfield  {journal} {\bibinfo  {journal} {Phys. Rev. B}\ }\textbf {\bibinfo
  {volume} {97}},\ \bibinfo {pages} {205127} (\bibinfo {year}
  {2018})}\BibitemShut {NoStop}%
\bibitem [{\citenamefont {Liang}\ and\ \citenamefont
  {Prendergast}(2019)}]{PhysRevB.100.075121}%
  \BibitemOpen
  \bibfield  {author} {\bibinfo {author} {\bibfnamefont {Y.}~\bibnamefont
  {Liang}}\ and\ \bibinfo {author} {\bibfnamefont {D.}~\bibnamefont
  {Prendergast}},\ }\href {https://doi.org/10.1103/PhysRevB.100.075121}
  {\bibfield  {journal} {\bibinfo  {journal} {Phys. Rev. B}\ }\textbf {\bibinfo
  {volume} {100}},\ \bibinfo {pages} {075121} (\bibinfo {year}
  {2019})}\BibitemShut {NoStop}%
\bibitem [{\citenamefont {Vu}\ \emph {et~al.}(2020)\citenamefont {Vu},
  \citenamefont {Lavrentyev}, \citenamefont {Gabrelian}, \citenamefont {Tkach},
  \citenamefont {Pham}, \citenamefont {Marchuk}, \citenamefont {Parasyuk},\
  and\ \citenamefont {Khyzhun}}]{VU2020106287}%
  \BibitemOpen
  \bibfield  {author} {\bibinfo {author} {\bibfnamefont {T.~V.}\ \bibnamefont
  {Vu}}, \bibinfo {author} {\bibfnamefont {A.}~\bibnamefont {Lavrentyev}},
  \bibinfo {author} {\bibfnamefont {B.}~\bibnamefont {Gabrelian}}, \bibinfo
  {author} {\bibfnamefont {V.}~\bibnamefont {Tkach}}, \bibinfo {author}
  {\bibfnamefont {K.~D.}\ \bibnamefont {Pham}}, \bibinfo {author}
  {\bibfnamefont {O.}~\bibnamefont {Marchuk}}, \bibinfo {author} {\bibfnamefont
  {O.}~\bibnamefont {Parasyuk}},\ and\ \bibinfo {author} {\bibfnamefont
  {O.}~\bibnamefont {Khyzhun}},\ }\href
  {https://doi.org/https://doi.org/10.1016/j.solidstatesciences.2020.106287}
  {\bibfield  {journal} {\bibinfo  {journal} {Solid State Sci.}\ }\textbf
  {\bibinfo {volume} {104}},\ \bibinfo {pages} {106287} (\bibinfo {year}
  {2020})}\BibitemShut {NoStop}%
\bibitem [{\citenamefont {Mortensen}\ \emph {et~al.}(2017)\citenamefont
  {Mortensen}, \citenamefont {Seidler}, \citenamefont {Kas}, \citenamefont
  {Govind}, \citenamefont {Schwartz}, \citenamefont {Pemmaraju},\ and\
  \citenamefont {Prendergast}}]{PhysRevB.96.125136}%
  \BibitemOpen
  \bibfield  {author} {\bibinfo {author} {\bibfnamefont {D.~R.}\ \bibnamefont
  {Mortensen}}, \bibinfo {author} {\bibfnamefont {G.~T.}\ \bibnamefont
  {Seidler}}, \bibinfo {author} {\bibfnamefont {J.~J.}\ \bibnamefont {Kas}},
  \bibinfo {author} {\bibfnamefont {N.}~\bibnamefont {Govind}}, \bibinfo
  {author} {\bibfnamefont {C.~P.}\ \bibnamefont {Schwartz}}, \bibinfo {author}
  {\bibfnamefont {S.}~\bibnamefont {Pemmaraju}},\ and\ \bibinfo {author}
  {\bibfnamefont {D.~G.}\ \bibnamefont {Prendergast}},\ }\href
  {https://doi.org/10.1103/PhysRevB.96.125136} {\bibfield  {journal} {\bibinfo
  {journal} {Phys. Rev. B}\ }\textbf {\bibinfo {volume} {96}},\ \bibinfo
  {pages} {125136} (\bibinfo {year} {2017})}\BibitemShut {NoStop}%
\bibitem [{\citenamefont {Hong}\ \emph {et~al.}(2015)\citenamefont {Hong},
  \citenamefont {Stoerzinger}, \citenamefont {Moritz}, \citenamefont
  {Devereaux}, \citenamefont {Yang},\ and\ \citenamefont
  {Shao-Horn}}]{doi:10.1021/jp511931y}%
  \BibitemOpen
  \bibfield  {author} {\bibinfo {author} {\bibfnamefont {W.~T.}\ \bibnamefont
  {Hong}}, \bibinfo {author} {\bibfnamefont {K.~A.}\ \bibnamefont
  {Stoerzinger}}, \bibinfo {author} {\bibfnamefont {B.}~\bibnamefont {Moritz}},
  \bibinfo {author} {\bibfnamefont {T.~P.}\ \bibnamefont {Devereaux}}, \bibinfo
  {author} {\bibfnamefont {W.}~\bibnamefont {Yang}},\ and\ \bibinfo {author}
  {\bibfnamefont {Y.}~\bibnamefont {Shao-Horn}},\ }\href
  {https://doi.org/10.1021/jp511931y} {\bibfield  {journal} {\bibinfo
  {journal} {J. Phys. Chem. C .}\ }\textbf {\bibinfo {volume} {119}},\ \bibinfo
  {pages} {2063} (\bibinfo {year} {2015})}\BibitemShut {NoStop}%
\bibitem [{\citenamefont {B.~Grad}\ \emph {et~al.}(2018)\citenamefont
  {B.~Grad}, \citenamefont {R.~González}, \citenamefont {Torres~Díaz},\ and\
  \citenamefont {V.~Bonzi}}]{JMSRR2018}%
  \BibitemOpen
  \bibfield  {author} {\bibinfo {author} {\bibfnamefont {G.}~\bibnamefont
  {B.~Grad}}, \bibinfo {author} {\bibfnamefont {E.}~\bibnamefont
  {R.~González}}, \bibinfo {author} {\bibfnamefont {J.}~\bibnamefont
  {Torres~Díaz}},\ and\ \bibinfo {author} {\bibfnamefont {E.}~\bibnamefont
  {V.~Bonzi}},\ }\href@noop {} {\bibfield  {journal} {\bibinfo  {journal}
  {Journal of Materials Science Research and Reviews}\ }\textbf {\bibinfo
  {volume} {1}},\ \bibinfo {pages} {1} (\bibinfo {year} {2018})}\BibitemShut
  {NoStop}%
\bibitem [{\citenamefont {Maitra}(2022)}]{maitra2021double}%
  \BibitemOpen
  \bibfield  {author} {\bibinfo {author} {\bibfnamefont {N.~T.}\ \bibnamefont
  {Maitra}},\ }\href {https://doi.org/10.1146/annurev-physchem-082720-124933}
  {\bibfield  {journal} {\bibinfo  {journal} {Annu. Rev. Phys. Chem}\ }\textbf
  {\bibinfo {volume} {73}},\ \bibinfo {pages} {null} (\bibinfo {year}
  {2022})}\BibitemShut {NoStop}%
\bibitem [{\citenamefont {Elliott}\ \emph {et~al.}(2011)\citenamefont
  {Elliott}, \citenamefont {Goldson}, \citenamefont {Canahui},\ and\
  \citenamefont {Maitra}}]{ELLIOTT2011110}%
  \BibitemOpen
  \bibfield  {author} {\bibinfo {author} {\bibfnamefont {P.}~\bibnamefont
  {Elliott}}, \bibinfo {author} {\bibfnamefont {S.}~\bibnamefont {Goldson}},
  \bibinfo {author} {\bibfnamefont {C.}~\bibnamefont {Canahui}},\ and\ \bibinfo
  {author} {\bibfnamefont {N.~T.}\ \bibnamefont {Maitra}},\ }\href
  {https://doi.org/https://doi.org/10.1016/j.chemphys.2011.03.020} {\bibfield
  {journal} {\bibinfo  {journal} {Chem. Phys.}\ }\textbf {\bibinfo {volume}
  {391}},\ \bibinfo {pages} {110} (\bibinfo {year} {2011})}\BibitemShut
  {NoStop}%
\bibitem [{\citenamefont {Lange}\ and\ \citenamefont
  {Aziz}(2013)}]{C3CS00008G}%
  \BibitemOpen
  \bibfield  {author} {\bibinfo {author} {\bibfnamefont {K.~M.}\ \bibnamefont
  {Lange}}\ and\ \bibinfo {author} {\bibfnamefont {E.~F.}\ \bibnamefont
  {Aziz}},\ }\href {https://doi.org/10.1039/C3CS00008G} {\bibfield  {journal}
  {\bibinfo  {journal} {Chem. Soc. Rev.}\ }\textbf {\bibinfo {volume} {42}},\
  \bibinfo {pages} {6840} (\bibinfo {year} {2013})}\BibitemShut {NoStop}%
\end{thebibliography}%

\clearpage

\newpage

\begin{widetext}
\renewcommand{\thefigure}{S\arabic{figure}}
\setcounter{figure}{0}
\renewcommand{\theequation}{S\arabic{equation}}
\setcounter{equation}{0}
\renewcommand{\thepage}{S\arabic{page}}
\setcounter{page}{1}

\section*{Supplementary Information}

Let us consider the general case of optical dipole transitions between many body states, initial and final, comprising single-particle orbitals from distinct self-consistent fields. The many-body transition operator can be written, with respect to the initial state, as:
\begin{align}
    \hat{O} = \sum_{i,j} \braket{\phi_i|\hat{o}|\phi_j} a_i^\dagger a_j + 
    \braket{\phi_j|\hat{o}|\phi_i} a_i a_j^\dagger,
\end{align}
where the ordering of operations is deliberate and the sum runs over the entire single-particle Hilbert space. These transitions may result from photoabsorption or may define photoemission, and are individually weighted by their single-particle dipole matrix elements: $o_{ij} = \braket{\phi_i|\hat{o}|\phi_j}$, where $\hat{o} = \bold{\epsilon \cdot \hat{r}}$ defines the electric field polarization and its interaction with the electronic position. Whether these transitions are allowed or not depends on the many-body state and its orbital occupations. We will work within the single Slater determinant approximation, where the $N$ occupied orbitals that define the self-consistent field via the electron density can also define a many-body state, emergent from the vacuum, as:
\begin{align}
    \ket{\Psi}=\prod_{i=1}^N a_i^\dagger \ket{0}
\end{align}
With this notation, we can define the many-body transition amplitude from an initial state $\ket{\Psi_I}$ to a final state $\ket{\Psi_F}$ as:
\begin{align}\label{M_IF}
    M^{I,F} = \sum_i^{\text{unocc}} \sum_j^{\text{occ}} \braket{\Psi_F | \Psi_I^{i-j}}
    o_{ij} ,
\end{align}
where the notation $\Psi_I^{i-j}$ implies the annihilation of the electron in orbital $j$ and the creation of an electron in orbital $i$ -- a non-self-consistent creation of an electron-hole pair within the orbitals defined by the initial state. The remaining overlap with the $N$-electron final state amounts also to an $N \times N$ Slater determinant defined by the overlap matrix between the occupied orbitals of the respective self-consistent fields:
\begin{align}\label{MB_overlap}
    \braket{\Psi_F | \Psi_I^{i-j}} = {}^F\Xi_{i-j}
\end{align}
where ${}^F\Xi_{i-j}$ is the determinant (with a factor of $\pm 1$) of an $N \times N$ matrix of orbital overlaps $\braket{\phi_k^F | \phi_l^I}$, between initial and final state orbitals, and the subscript $i-j$ indicates that column $j$ has been replaced by column $i$.

For the XES process, $\Psi_I$ is the state with a core-hole and can be approximated by the expression shown in Eq.~\ref{I_SD}, while $\Psi_F$ is a core-filled state with a valence hole. Due to the localized nature of the core electrons, it is reasonable to expect the core orbital of $\Psi_F$ to have negligible overlap with any electron present in $\Psi_I$. Consequently, for XES, $\braket{\Psi^F | \Psi_I^{i-j}} \approx 0$ unless $i = 0$, where $0$ denotes a core orbital. This leads to the following simplification of Eq.~\ref{M_IF}

\begin{align}
    M^{I,F} = \sum_j^{\text{occ}} \braket{\Psi_F | \Psi_I^{0-j}}
    o_{cj}.
\end{align}

Thus, for ooDFT (MBXES), which approximates each finals state $\ket{\Psi_F}$ with Eq.~\ref{StateSpecific_SD} (Eq.~\ref{MBXES_SD}), the transition element $M^{\rm{I,F}}$ reduces to Eq.~\ref{ESD_mat} (Eq.~\ref{MBXES_mat}). We can expect the core-orbital subspaces of $\ket{\Psi_F}$ and $\Psi_I^{0-j}$ to be approximately equal, for any $F$ and $j$. First-order perturbation theory would support this approximation, especially due to the energy isolation of  the  core  orbitals  from  the  valence  subspace - perturbations  scaling  as  the  inverse  of  orbital  energy  differences. Therefore, in the overlap term $\braket{\Psi_F | \Psi_I^{0-j}}$, the core-contribution, which can be approximated as unity, can be factored out and the overlap can be evaluated exclusively in terms of the valence orbitals. This is particularly useful in pseudopotential based calculations, where the valence KS orbitals are readily available. For a final-state $\ket{\Psi_F}$ having a hole in the $f$-th orbital, the many-body overlap of Eq.~\ref{MB_overlap} can then be written as

\begin{align}\label{OvlpMat}
    {}^F\Xi_{0-j} = (-1)^{f+j} {}^F{\Xi}'_{j,f}
\end{align}
such that
\begin{align}\label{Determinant_xiprime}
    {}^F{\Xi}'_{j,f} = \mathrm{det } \begin{bmatrix}
{{}^F\xi_{11}} & \hdots & {{}^F\xi_{1,f-1}} &{{}^F\xi_{1,f+1}} &  \hdots & {{}^F\xi_{1,N}}\\
\vdots & \ddots & \vdots & \vdots & \ddots & \vdots \\
{{}^F\xi_{j-1,1}} & \hdots & {{}^F\xi_{j-1,f-1}} &{{}^F\xi_{j-1,f+1}} &  \hdots & {{}^F\xi_{j-1,N}}\\
{{}^F\xi_{j+1,1}} & \hdots & {{}^F\xi_{j+1,f-1}} &{{}^F\xi_{j+1,f+1}} &  \hdots & {{}^F\xi_{j+1,N}}\\
\vdots & \ddots & \vdots & \vdots & \ddots & \vdots \\
{{}^F\xi_{N,1}} & \hdots & {{}^F\xi_{N,f-1}} &{{}^F\xi_{N,f+1}} &  \hdots & {{}^F\xi_{N,N}}.\\
\end{bmatrix}
\end{align}
where ${}^F\xi_{p,q} = \braket{\phi^F_p|\phi^I_q}$ denotes the single-particle overlap between the $p$-th final-state and the $q$-th initial-state orbital. 

Note that in the main text, the term ${}^F\Xi_{0-j}$ in the ooDFT (MBXES) approximation has been denoted by ${}^{(f)}\Xi_{j,f}$ ($\Xi_{j,f}$). Within the MBXES approximation, which constructs each $\ket{\Psi_F}$ from the frozen ground-state KS orbitals, the single-particle overlaps are independent of $F$ and can be written as ${}^F\xi_{p,q} = \xi_{p,q}$.
\subsection*{Projected Ground State Single Particle Formalism}
The MBXES transition dipole moment is 
\begin{align*}
   M^f_{\rm{MBXES}}= \sum_{j=1}^N  \Xi_{j,f} o_{jf}
\end{align*}
In the pGS treatment, the many-electron overlap $\Xi_{j,f}$ is approximated as follows
\begin{align}
        \Xi_{j,f}  = \braket{\Psi_{-f}|\Psi_{\mathrm{I}}^{+0-j}} =\braket{\Psi_{\rm{GS}}|a_f^\dagger \tilde{a}_j|\Psi_{\rm{I}}^{+0}} = - \braket{\Psi_{\rm{GS}}|\tilde{a}_j a_f^\dagger|\Psi_{\rm{I}}^{+0}} \approx -\braket{\tilde{\phi_j}|{\phi}_f} = -\xi^*_{j,f}.
\end{align}
Then, ignoring the constant phase-factor $-1$, the transition dipole moment becomes
\begin{align*}
    M^f_{\rm{pGS}} = \sum_{j=1}^N \xi_{j,f}^* o_{jf},
\end{align*}
which is the expression in Eq.~\ref{pGS_Mat}.
\subsection*{Relation between $M^i_{\rm{MBXES}}$ and $M^i_{\rm{GS}}$}
From Eq.~\ref{OvlpMat}, we have  
\begin{align}\label{MBXES_mat2}
    M_{\rm{MBXES}}^f = \sum_{j=1}^N (-1)^{f+j} \Xi'_{j,f} \braket{\phi_0|\hat{o}|\tilde{\phi}_j}.
\end{align}
Now, if $C$ is the cofactor of the overlap matrix $\xi$, then
\begin{align}
    \xi^{-1} = \frac{\mathrm{adj}(\xi)}{\mathrm{det}(\xi)} = \frac{C^T}{\mathrm{det}(\xi)}.
\end{align}
So,
\begin{align}
    \left(\xi^{-1}\right)_{j,f} = \frac{C_{f,j}}{\mathrm{det}(\xi)} = \frac{(-1)^{j+f}\left(\Xi'\right)_{f,j}}{\mathrm{det}(\xi)}
\end{align}
Therefore, if the $(N \times N)$ matrix $\xi$ is orthogonal, then
\begin{align}
    \xi_{j,f} = \pm 1 . (-1)^{j+f} \Xi'_{j,i},
\end{align}
and consequently,
\begin{align}
    |M^f_{\rm{MBXES}}|^2 = |M^f_{\rm{GS}}|^2.
\end{align}
So, if the electrons occupy the same space in the ground and the core-excited state, then the GS and the MBXES spectra will be identical.
\subsection*{De-excitation within the valence subspace}
Fig.~\ref{ValDeEx} shows the spectrum of de-excitation of phenol from an initial state with a hole in the 11-th KS orbital. Due to the low excitation energy (or, from an alternative perspective, due to the delocalized nature of the valence orbitals), the LR-TDDFT result, which is shown to be in good agreement with the ooDFT counterpart, is expected to be more reliable for such simulations. Additionally, the striking similarity between the many-body and the single-particle spectra in panel (b) can be attributed to the very small change in the occupied subspace (excepting the 11-th orbital) between $\ket{\Psi_{\rm{GS}}}$ and $\ket{\Psi_I}$ evident from a small (0.17 Debye) change in the corresponding net dipole-moment. This can be contrasted with the significantly larger change (3.01 Debye) associated with de-excitation to the oxygen core orbital, which gives rise to the noticeable difference between the MBXES and the GS spectra shown in Fig.~\ref{PhenolSpectra}(b). 
\begin{figure}[!htb]
\centering
\includegraphics[width=0.99\textwidth]{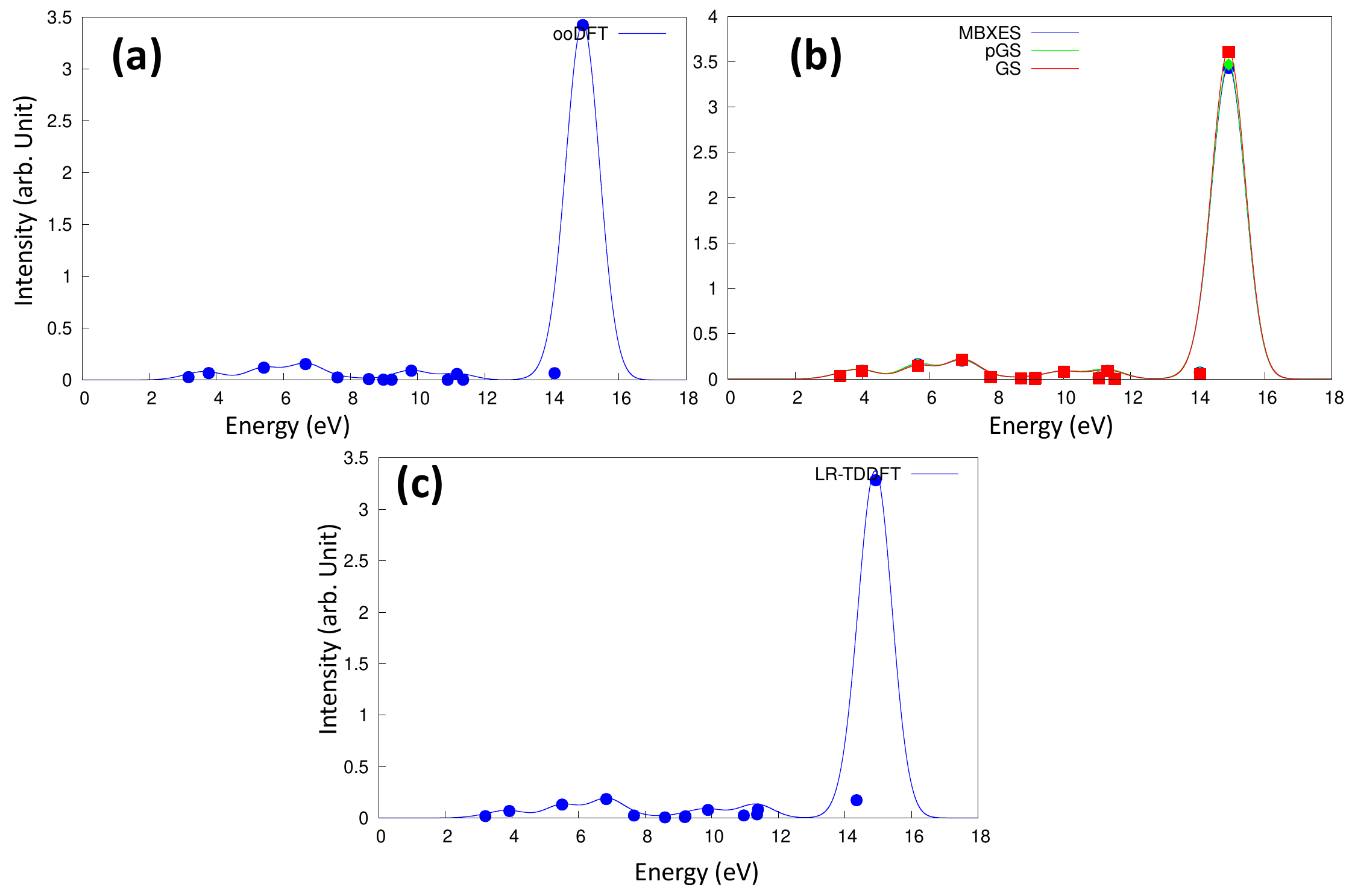}
\caption{Emission spectrum corresponding to electronic de-excitation of phenol from an initial state with a hole in the 11-th KS orbital. Panels (a) and (b) display the ooDFT and the LR-TDDFT results, respectively. The agreement between the two spectra can be attributed to the fact that, unlike the core-hole present in the $\ket{\Psi_{\rm{IN}}}$ of XES, the valence hole present in the initial state here is a relatively delocalized object. Creation of this valence hole is associated with a relatively small change in the net density and dipole-moment (0.17 Debye) of the other electrons.}\label{ValDeEx}
\end{figure}

\begin{figure}[!htb]
\centering
\includegraphics[width=0.8\textwidth]{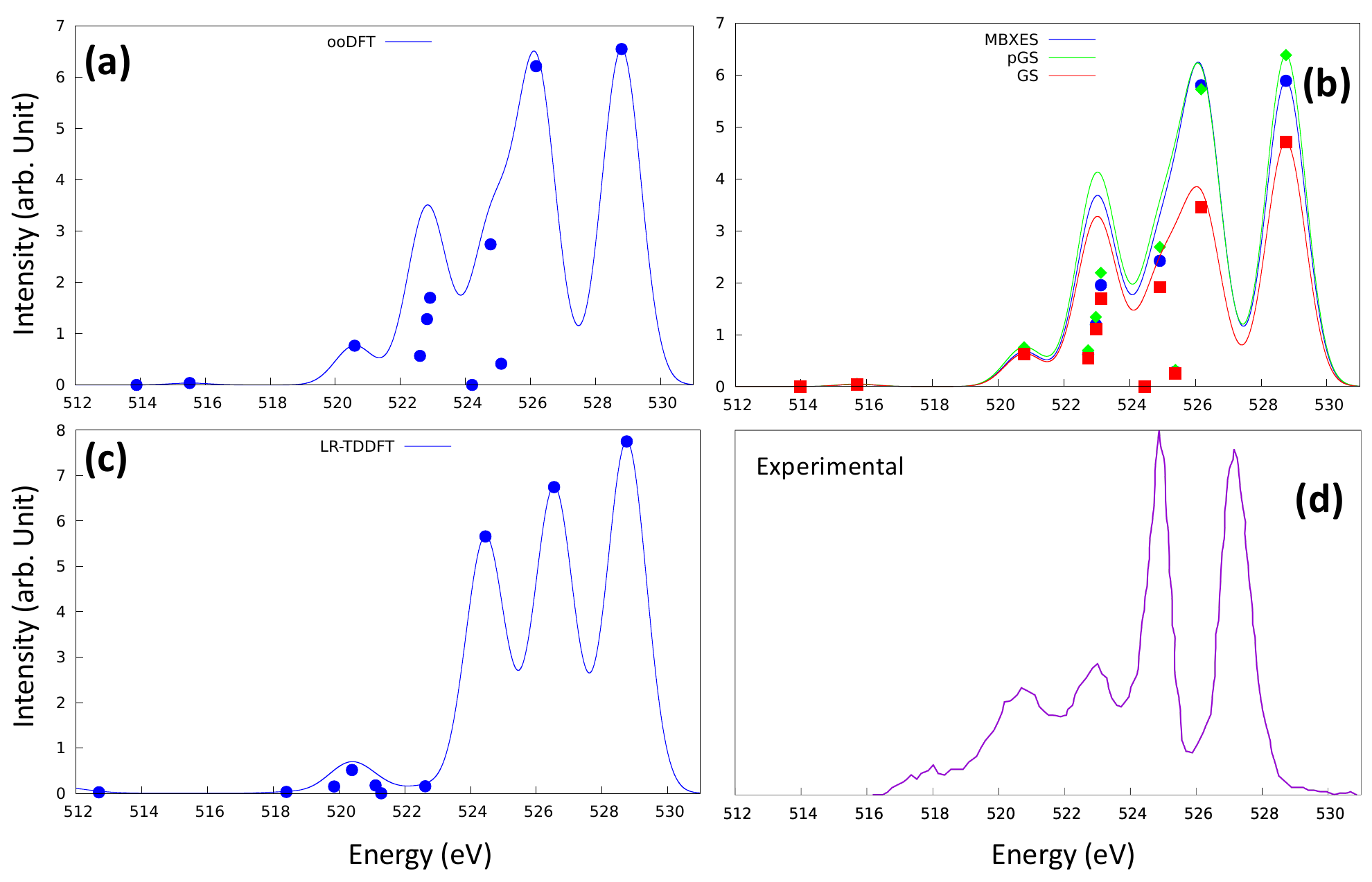}
\caption{Simulated and experimental~\cite{C3CS00008G} O K-edge emission spectra of an acetone molecule. $\omega$B97M-V xc-functional has been used in the calculations.}\label{AcetoneXES}
\end{figure}

\begin{figure}[!htb]
\centering
\includegraphics[width=0.8\textwidth]{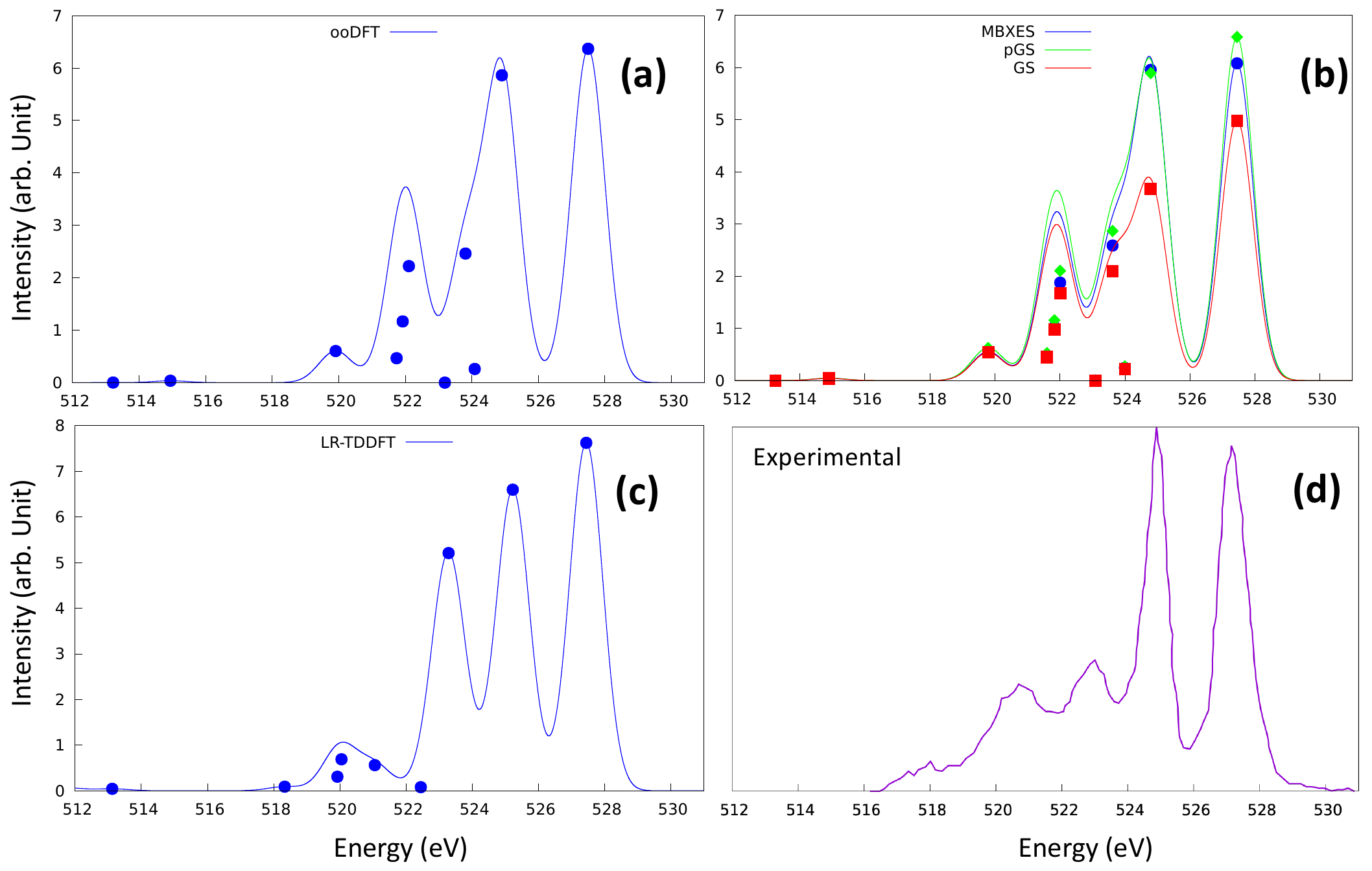}
\caption{Simulated and experimental~\cite{C3CS00008G} O K-edge emission spectra of an acetone molecule. B3LYP xc-functional has been used in the calculations.}\label{AcetoneXESB3LYP}
\end{figure}

\begin{figure}[!htb]
\centering
\includegraphics[width=0.8\textwidth]{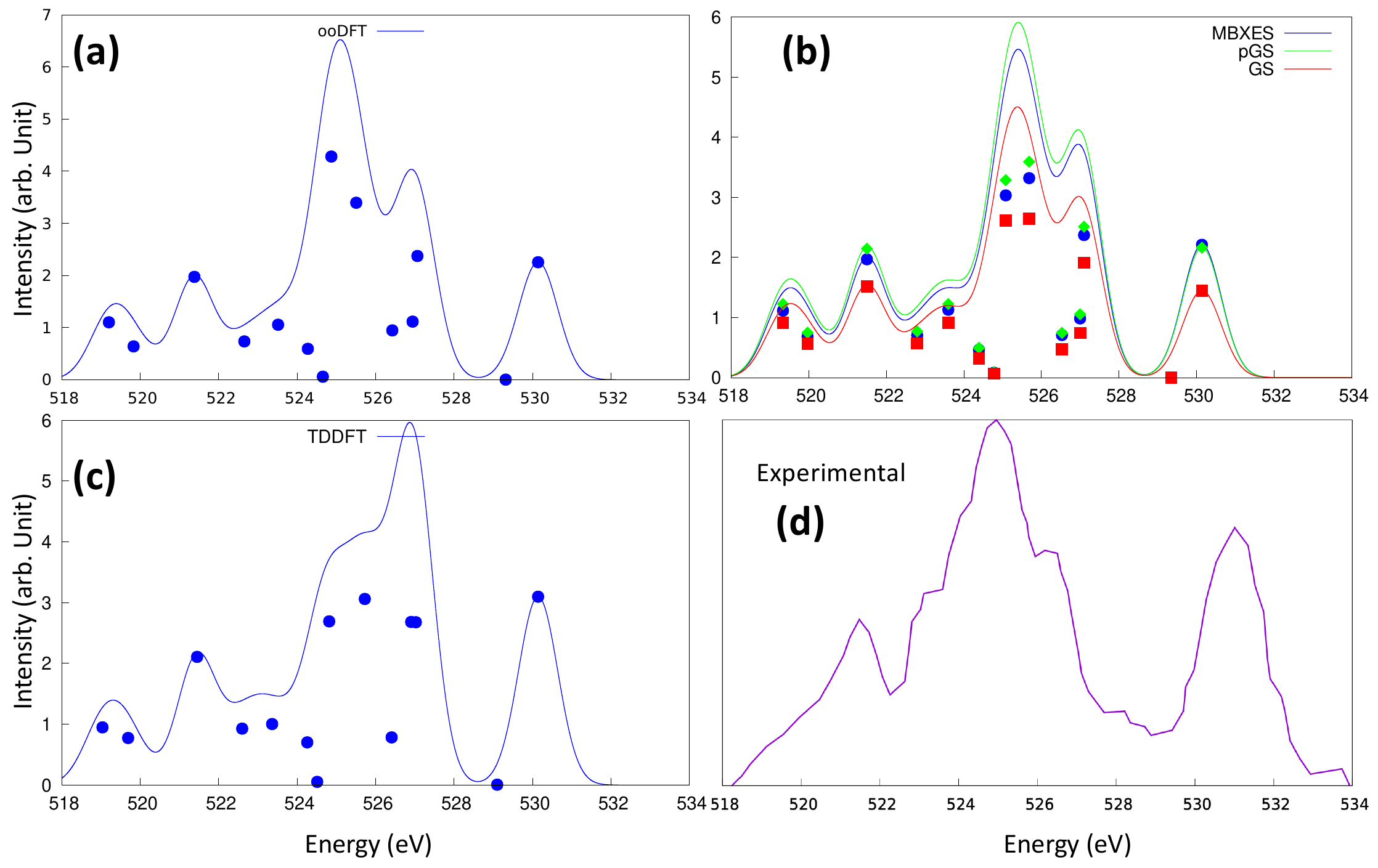}
\caption{Simulated and experimental~\cite{Yumatov1997} O K-edge emission spectra of a phenol molecule. B3LYP xc-functional has been used in the calculations.}\label{PhenolXESB3LYP}
\end{figure}

\begin{figure}[!htb]
\centering
\includegraphics[width=0.8\textwidth]{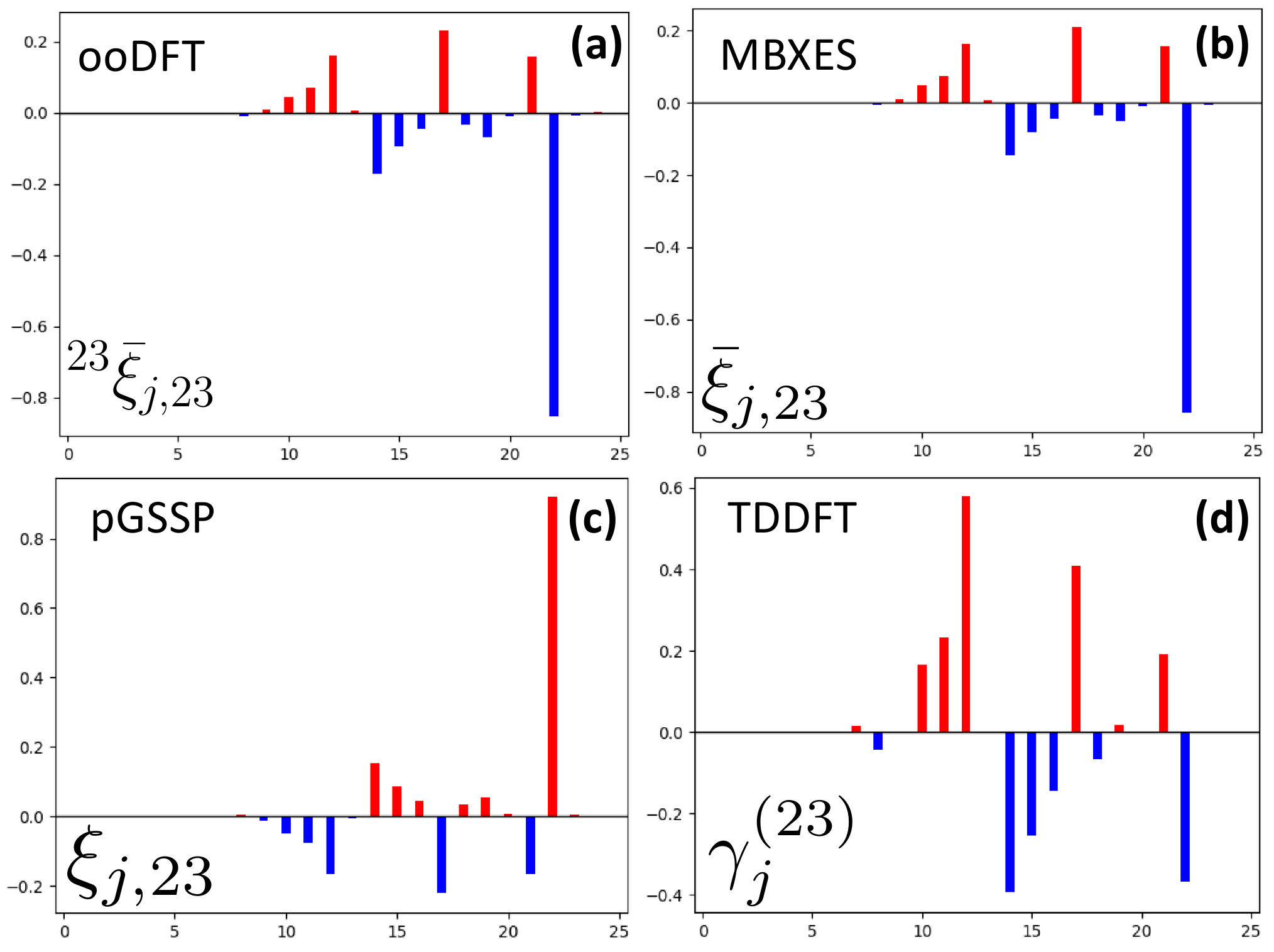}
\caption{Coefficients of $\braket{\phi_c|\hat{o}|\tilde{\phi_j}}$ corresponding to the 23rd de-excitation, as a function of $j$, for the different theoretical approaches presented in this paper. The ooDFT coefficient $\bar{{}^{(23)}\xi}_{j,23}$, the MBXES coefficient $\bar{\xi}_{j,23}$, the pGS coefficient $\xi_{j,23}$, and the LR-TDDFT coefficient $\gamma_j^{(23)}$ are shown in the bar charts in the panels (a), (b), (c) and (d), respectively. Red (blue) bars denote positive (negative) values. Note that an overall phase-factor of $\pm 1$ does not affect the oscillator strength.}\label{BarPlot}
\end{figure}

\begin{figure}[!htb]
\centering
\includegraphics[width=0.99\textwidth]{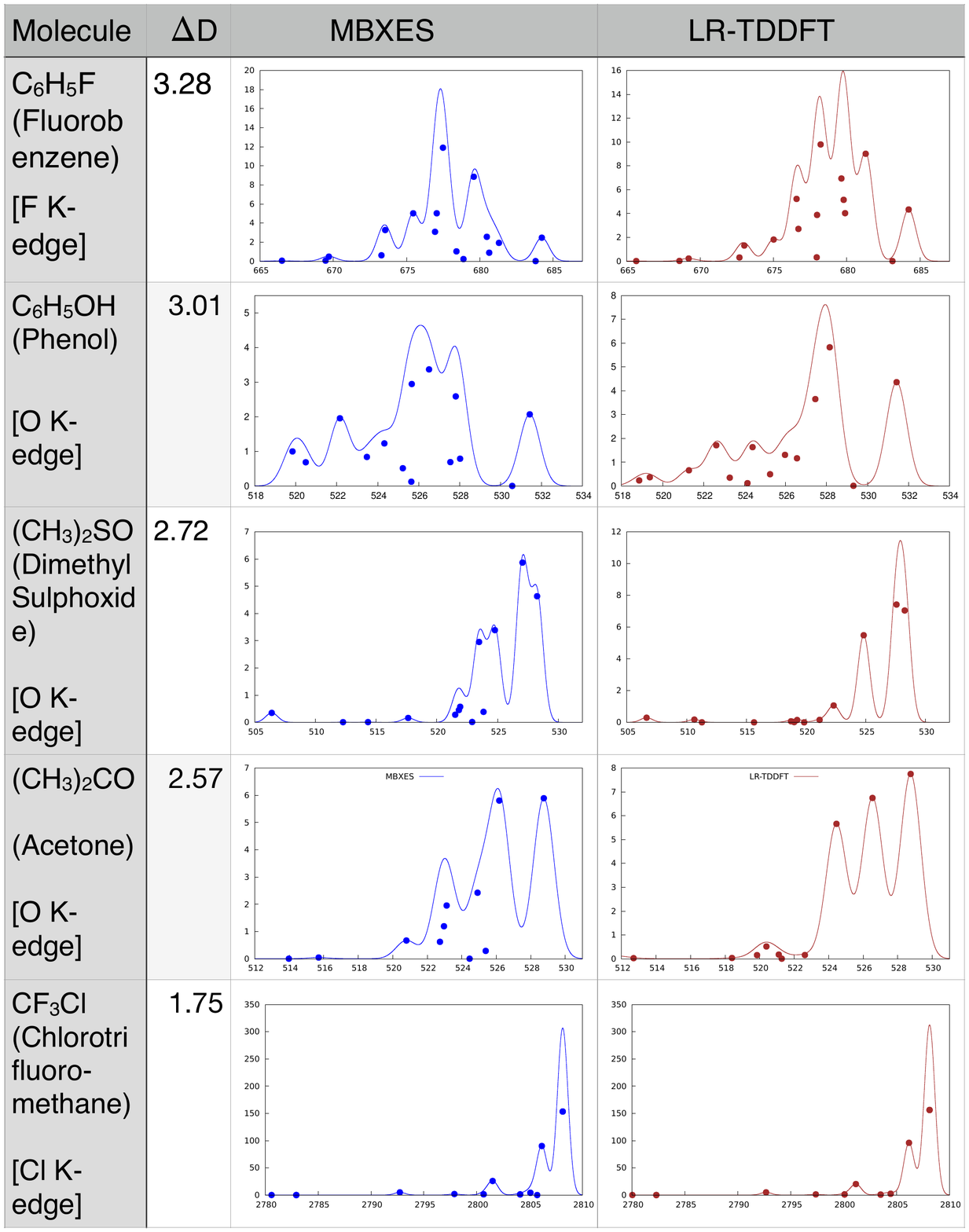}
\caption{Table showing simulated MBXES and LR-TDDFT spectra of different molecules. The second column shows the change in the net dipole moment, in Debye units, of the remaining electrons following the relevant core-ionization. In each plot, the horizontal (vertical) axis corresponds to energy of emitted photon (intensity of emission) in eV (arbitrary unit).}\label{Table_1}
\end{figure}

\begin{figure}[!htb]
\centering
\includegraphics[width=0.99\textwidth]{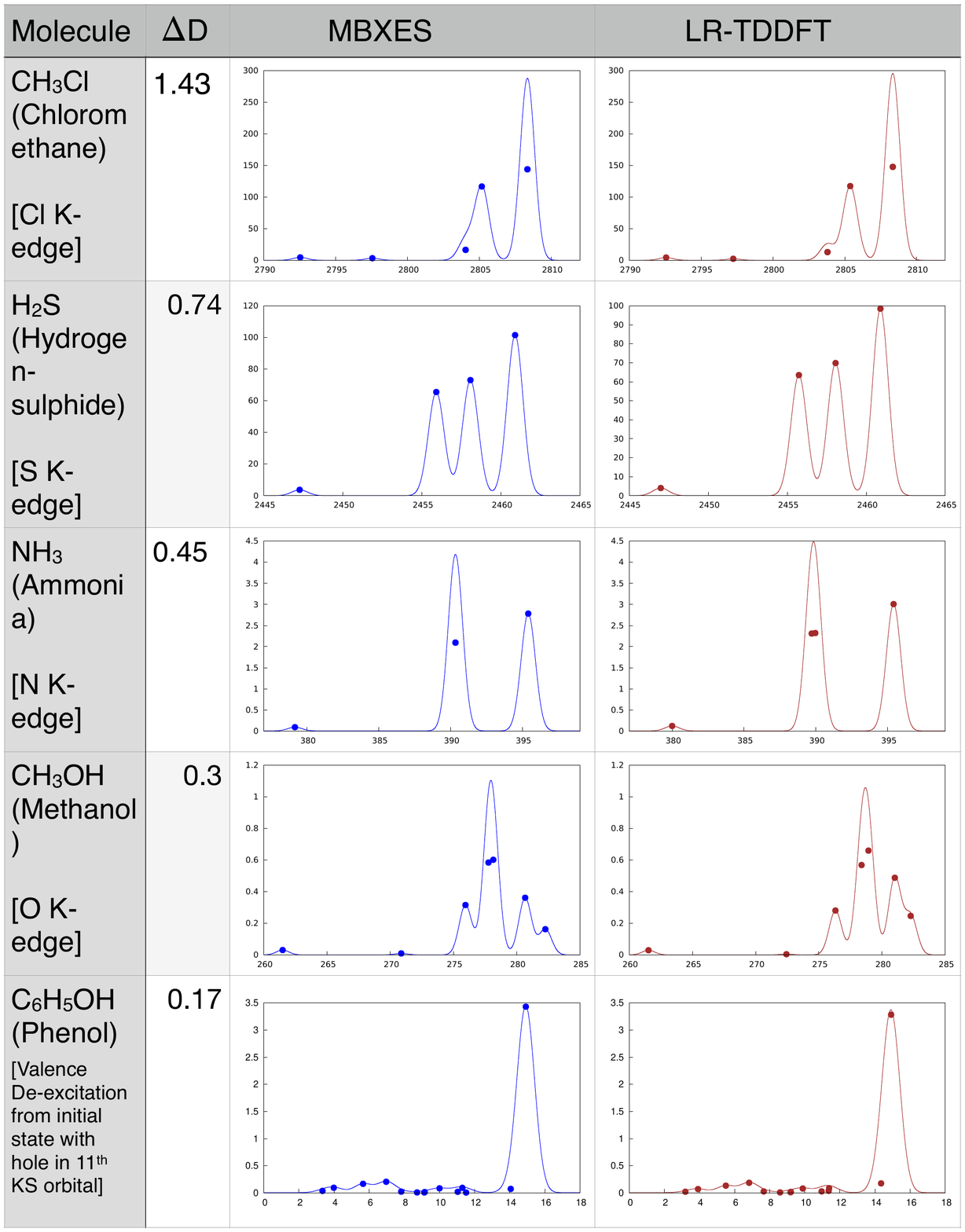}
\caption{Table showing simulated MBXES and LR-TDDFT spectra of different molecules. The second column corresponds to the change in the net dipole moment, in Debye units, of the remaining electrons following the relevant ionization. In each plot, the horizontal (vertical) axis corresponds to energy of emitted photon (intensity of emission) in eV (arbitrary unit).}\label{Table_2}
\end{figure}

\begin{figure}[!htb]
\centering
\includegraphics[width=0.99\textwidth]{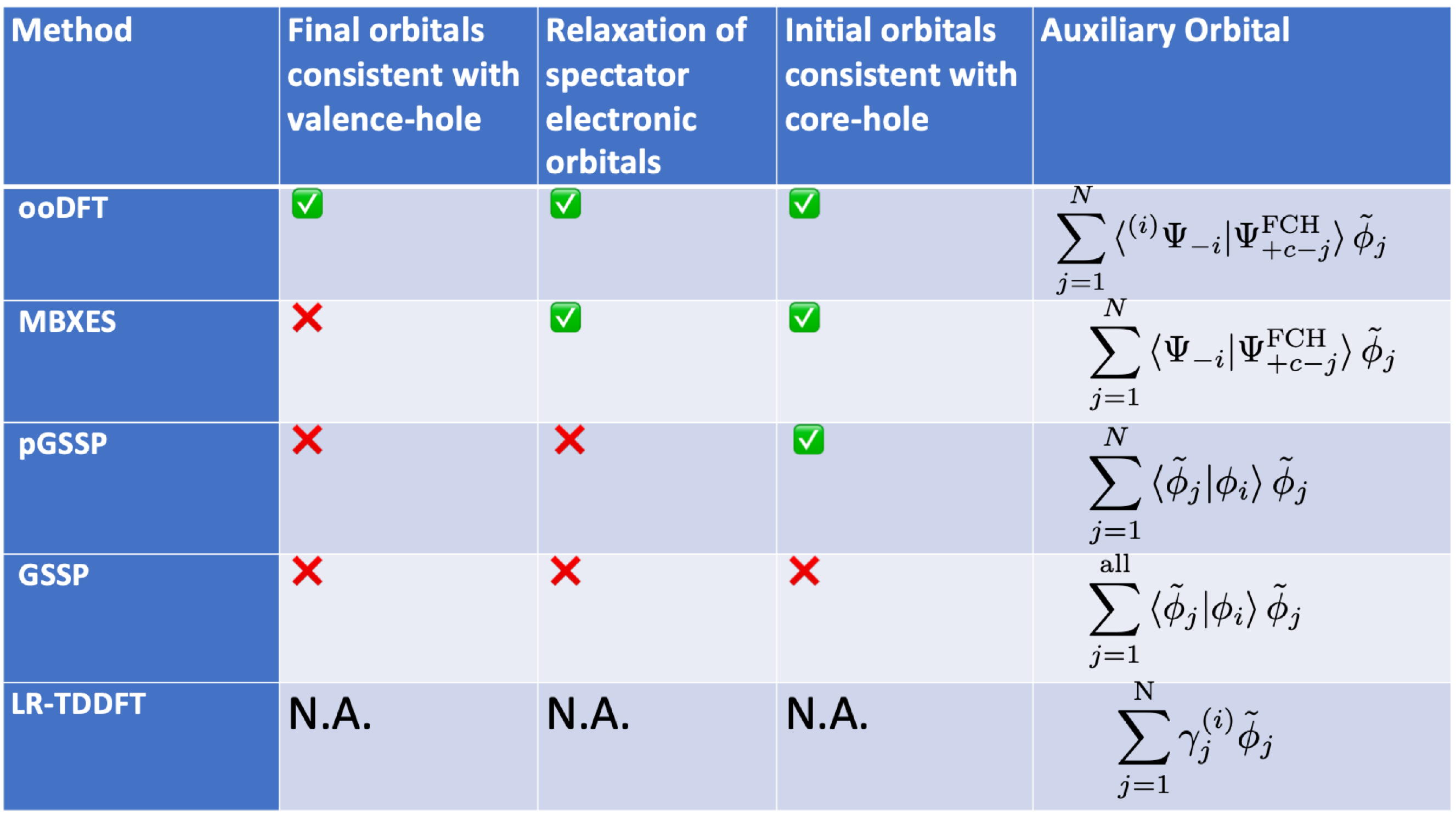}
\caption{Table summarizing important properties and approximations corresponding to various simulation-methods discussed in the paper. Note that, in each case, the transition-dipole moment is given by $\braket{\textrm{core}|\hat{o}|\textrm{Auxiliary Orbital}}$, where $\hat{o}$ is the single-particle dipole operator.}\label{Table}
\end{figure}

\end{widetext}

\end{document}